\documentclass[journal,12pt,onecolumn,draftclsnofoot,]{IEEEtran}
\usepackage{floatrow}
\usepackage{graphicx}
\usepackage[labelseparator=none]{subfig}
\usepackage{caption}
\usepackage{amsmath,amsfonts}
\usepackage{amssymb}
\usepackage{xcolor}
\usepackage{algorithmic}
\usepackage{algorithm}
\usepackage{array}
\usepackage{textcomp}
\usepackage{url}
\usepackage{verbatim}
\usepackage{cite}
\usepackage{hyperref}
\usepackage[capitalize]{cleveref}
\allowdisplaybreaks[4]
\begin{document}

\title{SAFARI: Sparsity enabled Federated Learning with Limited and Unreliable Communications}

\author{Yuzhu Mao$^{*}$, Zihao Zhao$^{*}$, Meilin Yang, Le Liang, Yang Liu, Wenbo Ding,\\Tian Lan,~\IEEEmembership{Senior Member,~IEEE}, Xiao-Ping Zhang,~\IEEEmembership{Fellow,~IEEE}
\thanks{$^{*}$ These authors contribute equally.}
\thanks{Y.~Mao, Z.~Zhao, M.~Yang, W.~Ding, and X.-P.~Zhang are with Tsinghua-Berkeley Shenzhen Institute, Tsinghua Shenzhen International Graduate School, Tsinghua University, China. W.~Ding is the corresponding author. E-mail: (\{myz20, zhao-zh21,yml21\} @mails.tsinghua.edu.cn, ding.wenbo@sz.tsinghua.edu.cn)}
\thanks{W.~Ding is also with RISC-V International Open Source Laboratory, Shenzhen, China, 518055.}
\thanks{L.~Liang is with the National Mobile Communications Research Laboratory, Southeast University, Nanjing 210096, and also with the Purple Mountain Laboratories, Nanjing 211111, China. E-mail: (lliang@seu.edu.cn)}
\thanks{Y.~Liu is with the Institute for AI Industry Research (AIR), Tsinghua University, China. E-mail: (liuy03@air.tsinghua.edu.cn)}
\thanks{T.~Lan is with the Department of Electrical and Computer Engineering, George Washington University, DC, USA. Email: (tlan@gwu.edu)}
\thanks{X.-P.~Zhang is also with the Department of Electrical, Computer and Biomedical Engineering, Ryerson University, Toronto, ON M5B 2K3, Canada. E-mail: (xzhang@ee.ryerson.ca)}
}



\maketitle

\begin{abstract}
Federated learning (FL) enables edge devices to collaboratively learn a model in a distributed fashion. Many existing researches have focused on improving communication efficiency of high-dimensional models and addressing bias caused by local updates. However, most of FL algorithms are either based on reliable communications or assume fixed and known unreliability characteristics. In practice, networks could suffer from dynamic channel conditions and non-deterministic disruptions, with time-varying and unknown characteristics. To this end, in this paper we propose a sparsity enabled FL framework with both communication efficiency and bias reduction, termed as SAFARI. It makes novel use of a similarity among client models to rectify and compensate for bias that is resulted from unreliable communications. More precisely, sparse learning is implemented on local clients to mitigate communication overhead, while to cope with unreliable communications, a similarity-based compensation method is proposed to provide surrogates for missing model updates. We analyze SAFARI under bounded dissimilarity and with respect to sparse models. It is demonstrated that SAFARI under unreliable communications is guaranteed to converge at the same rate as the standard FedAvg with perfect communications. Implementations and evaluations on CIFAR-10 dataset validate the effectiveness of SAFARI by showing that it can achieve the same convergence speed and accuracy as FedAvg with perfect communications, with up to 80\% of the model weights being pruned and a high percentage of client updates missing in each round. 
\end{abstract}

\section{Introduction}


With rapid deployment of mobile sensing and computing devices, there are growing interests in fully exploiting distributed computing resources, as well as huge volumes of data generated at network edge, for efficient learning~\cite{qin2021federated}. To this end, federated learning (FL)~\cite{mcmahan2017communication,Yang-et-al:2019} enables distributed edge devices to collaboratively learn a model while maintaining data privacy~\cite{DifferentialPrivacy,GoogleSecure,liu2020secure}, by allowing a central server and distributed clients to exchange updated model parameters and performing global aggregations. As wireless communications in practice often have limited network capacity~\cite{mcmahan2017communication, qin2021federated}, a number of proposals have been made on the communication-efficient FL. Examples include model pruning and sparsity to exploit the structural redundancy of dense models~\cite{ma2021effective} and leveraging multiple local training epochs before periodical global aggregation in order to mitigate communication overhead ~\cite{woodworth2020local, koloskova2020unified, khaled2020tighter}.


Nevertheless, most of existing FL algorithms either are based on reliable communications~\cite{koloskova2020unified, khaled2020tighter} or assume fixed and known unreliability characteristics~\cite{ye2022decentralized, fraboni2021clustered}. These assumptions may not hold in real-world FL applications. Protocols for data-intensive communications like the lightweight User Datagram Protocol (UDP) tend to focus on best effort delivery without mechanisms for detecting failures and re-transmission. Reliable transmission of local updates cannot be guaranteed~\cite{ye2022decentralized}. Further, an underlying wireless network could suffer from dynamic channel conditions and non-deterministic disruptions, whose characteristics are often unknown and time-varying. 
This raises serious challenges in FL -- unpredictable absences of local updates with time-varying characteristics would lead to non-homogeneous bias under non-IID data distribution, potentially introducing an unknown drift and causing slow and unstable convergence.


In this paper, we propose a Sparsity enAbled Federated leArning framework under limited and unReliable communIcations, termed as SAFARI. When unreliability characteristics are unknown and potentially time-varying, we show that it is possible to rectify the resulting bias in global model aggregation by leveraging similarity among different client models. More precisely, once distributed clients locally train their models with sparse algorithms, the central server (i) updates a similarity matrix tracking the similarity among different clients based on received sparse models, and (ii) for any absent update in the current round, substitutes it with an available update received from the most similar client. Intuitively, these similarity-based surrogates provide an optimal way of compensating for any missing local updates on the fly. This compensation works even if sparse algorithms are employed, as we show that similarity properties are preserved under sparsity. We formally analyze the impact of such compensations in FL and prove that under bounded dissimilarity (i.e., the difference among sparse models produced by different clients are bounded) and a sufficiently small learning rate, the proposed SAFARI algorithm is guaranteed to converge. Extensive evaluations over several popular sparse algorithms (including MAG and Synflow~\cite{tanaka2020pruning}) are conducted. The experiment results validate our theoretical analysis showing that the proposed SAFARI algorithm under unreliable communications achieves the same asymptotic convergence rate as standard FedAvg with reliable communications, even if 80\% of the model weights are pruned and a large percentage (up to 70\%) of client updates are lost in each round. SAFARI consistently achieves faster convergence than that without compensation under unreliable communications. 


The contributions of this paper are summarized as follows. 
\begin{itemize}
\item A sparsity enabled robust FL framework, SAFARI, is proposed to simultaneously reduce communication overhead and cope with unreliable communications in FL.
\item SAFARI leverages a novel similarity-based compensation scheme that actively tracks client similarity and substitutes any missing update (due to unreliable communications) on the fly with an available model update received from the most similar client.
\item We theoretically analyze the impact of such compensation with respect to sparse algorithms and prove that similarity properties are preserved under the use of sparse models. 
\item 
We establish global convergence analysis for SAFARI and demonstrate that even with limited and unreliable communications, SAFARI can achieve the same convergence rate of vanilla FedAvg with perfectly reliable communications. 
\item Experiments on CIFAR-10 dataset validate our theoretical analysis. SAFARI demonstrates fast and stable convergence under  unreliable communications and outperforms baselines without compensation. 
\end{itemize}

The rest of this paper is organized as follows. Section II introduces the background and related work as well as the motivation. In Section III, the proposed method is described in details. The theoretical analysis and the experimental results are provided in Section IV and V, respectively. Finally, conclusion remarks are summarized in Section VI.

\section{Background and Related work}
\subsection{Federated Learning}
Assume an FL system with one central server and $m$ distributed clients. Each client $i$ in the client set $\mathbb{M} = \{1, \dots, m\}$ has a local dataset $\mathit{D}_{i}$ of $n_i$ data samples. The goal of federated training is to solve the following optimization problem:  
\begin{align}
    \min_{\boldsymbol{x}\in \mathbb{R}^{d}} \mathcal{L}(\boldsymbol{x}) =\frac{1}{m}\sum_{i=1}^{m}\mathcal{L}_{i}(\boldsymbol{x}),
\label{optimization target}
\end{align}
where $\mathcal{L}_{i}(\boldsymbol{x}) = \sum_{i=1}^{m}\!\frac{1}{n_{i}}\!\sum_{z \in \mathit{D}_{i}}\!\ell_i(\boldsymbol{x}, z)$ is the local
objective function at the $i$-th client. Specifically, $z$ represents a data sample from $\mathit{D}_{i}$ and $\ell_{i}:\mathbb{R}^{d}\rightarrow\mathbb{R}$ is the local loss function based on the learning model $\boldsymbol{x}$ and client $i$'s own data.

In the $t$-th communication round, the server first broadcasts the global model $\boldsymbol{x}^{t}$ to clients. Then each client independently runs $\tau$ local iterations by optimization solver such as the stochastic gradient descent (SGD) from the current global model $\boldsymbol{x}^{t}$ to optimize its own local objective function  $\mathcal{L}_{i}(\boldsymbol{x})$. Take the SGD for example and the local iterations are as follows,
\begin{equation}
\begin{cases}
\boldsymbol{x}^{t}_{i, 0} = \boldsymbol{x}^{t}, \\ 
\boldsymbol{x}^{t}_{i, k} = \boldsymbol{x}^{t}_{i, k-1}-\frac{\eta}{\tau}g_{i}(\boldsymbol{x}^{t}_{i, k-1} |  \xi_{i,k}), k\in \mathbb{K},
\end{cases}
\end{equation}
where $\eta$ is the learning rate, $g_{i}(\boldsymbol{x}^{t}_{i, k-1}  |  \xi_{i,k})$ is the stochastic gradient computed with the data batch $\xi_{i,k}\sim \mathit{D}_{i}$, $\boldsymbol{x}^{t}_{i, k}$ is the local model after $k$ local iterations and $\mathbb{K}=\{1,\cdots,\tau\}$.

After completing $\tau$ iterations of local training, each client $i$ will send the new model $\boldsymbol{x}_{i, \tau}^{t}$ back to the central server, and the server will aggregate the received client models to update the global model by:
\begin{align}
\boldsymbol{x} ^{t+1} =  \frac{1}{m}\sum_{i=1}^{m}\boldsymbol{x}_{i, \tau}^{t}.
\label{eq:model without lazy}
\end{align} 

\subsection{Practical Issues and Related Work}
In the design and application of the FL system, there are some practical issues needed to be considered. According to a recent survey~\cite{qin2021federated}, the major issues in FL are summarized as SGD, robust aggregation, upload frequency, privacy leakage and wireless communications. Among these five issues, robust aggregation, upload frequency, and wireless communications are all related to the capability and reliability of communications, and the bias caused by consecutive local SGD steps cannot be neglected in all scenarios. Since privacy leakage is a separate line of research subjects that can always be combined with other works, in this paper we mainly focus on previous works that aim to address limited communication resources, unreliable communication links and local bias.

\textbf{Limited Communication Resources.} Edge devices in wireless networks usually have limited resources, especially for frequent communications. To reduce the transmission burden at each communication round, gradient or model weight compression is a mainstream technique, including quantization and sparsification. Gradient quantization maps each real-valued gradient/model element to a constant number of bits with lower-precision~\cite{alistarh2017qsgd, jhunjhunwala2021adaptive, bernstein2018signsgd}. As another line of work, sparsification prunes the dense gradient/model with a large amount of non-zero elements to a sparser one. In practice, these two compression techniques can be jointly used, and sparsification is usually the first step to reduce the number of weights for further quantization and transmission. The simplest way to sparsify a model is to keep only the coordinates with large magnitudes exceeding a selected threshold~\cite{strom2015scalable}. More sophisticated methods like unbiased sparsification and variance-reduced sparsified SGD have also been developed for training in a distributed fashion~\cite{wangni2018gradient,stich2018sparsified,karimireddy2019error}. One remaining question is that such sparsification operates after the local training completes, which provides no reduction on the computation and memory cost during training. 

As the training model becomes larger along with the growth of training data in recent years, sparse learning that pre-conducts sparsification and maintains sparse structure throughout training has been intensively investigated. In~\cite{mocanu2018scalable}, fully-connected layers were replaced by sparse ones achieved from an initial sparse topology with evolutionary algorithm before training. The connection sensitive had been investigated in~\cite{lee2018snip} for Single-Shot Network Pruning (SNIP). In~\cite{dettmers2019sparse}, the exponentially smoothed gradients was utilized to identify model layers and weights which reduced the error efficiently. You \textit{et al.} proposed to use the change of mask distances between epochs to identify a small sub-network at the early training stage, which could restore the comparable test accuracy to the dense network when being trained independently~\cite{you2019drawing}. Moreover, the sparse topology's updates based on parameter magnitudes and infrequent gradient calculations in~\cite{evci2020rigging} loosened the limitation on the size relationship between sparse model and the corresponding dense model, which further reduced the computation cost for sparse learning. However, despite the success empirical performance of the above sparse learning methods, theoretical analysis of the sparse model's property is still limited. 

\textbf{Unreliable Communications and Local Bias.} Due to the limited capability of distributed clients and communication channels, the communication reliability cannot be guaranteed in the FL system, especially with wireless networks~\cite{qin2021federated}. Previous work has proposed to address unreliable issues by optimizing the aggregation weights according to the link reliability matrix of communication links~\cite{ye2022decentralized}. Thus, it requires the knowledge of reliability matrix in advance, which is sometimes infeasible in real-world systems. 

To tackle the bias caused by local training steps, methods like drift-reduced SCAFFOLD~\cite{karimireddy2020scaffold} and Inexact DANE~\cite{reddi2016aide} with local approximate sub-solver have proved to be effective when the heterogeneity of local objectives is small enough. Recently, the Bias-Variance Reduced Local SGD algorithm surpasses non-local methods under a more relaxed second-order heterogeneity assumption~\cite{murata2021bias}. But the existing bias-reduction techniques still rely on reliable communications and do not consider the client bias caused by the random update loss. 

\subsection{Motivating Applications}
In this section, we provide several examples to explain some useful properties in practical FL systems that can be utilized to address the aforementioned issues.

\textbf{Local Computing Resources.} In FL collaborative systems, local clients are always equipped with a certain degree of computing power, no matter small edge devices as smartphones, werables and sensors, or distributed medical/financial institutes. It makes local sparse learning feasible, and reveals great potentials to achieve highly efficient local training with limited distributed resources.

\textbf{Clusterable Clients}. Although the non-IID data distribution and unstable clients of large amount remain challenging in FL systems, it is useful to notice that the clients in quite a few real-word systems tend to be clusterable in terms of data distribution. For example, in an Internet of vehicles system, vehicles within a certain area tend to record similar transportation information. Besides, the devices within the same smart home system usually collect the features of the same person. In these examples, the dissimilarity between client data in a certain group may be negligible, or even follow IID data distribution for the same learning task.

Note that although the pervasiveness of clusterable clients is demonstrated, the following analysis of our method is built upon the standard assumption on data dissimilarity as previous works~\cite{wang2020tackling}. 

\section{Methods}
\begin{figure}
    \centering
    \includegraphics[width=1\linewidth]{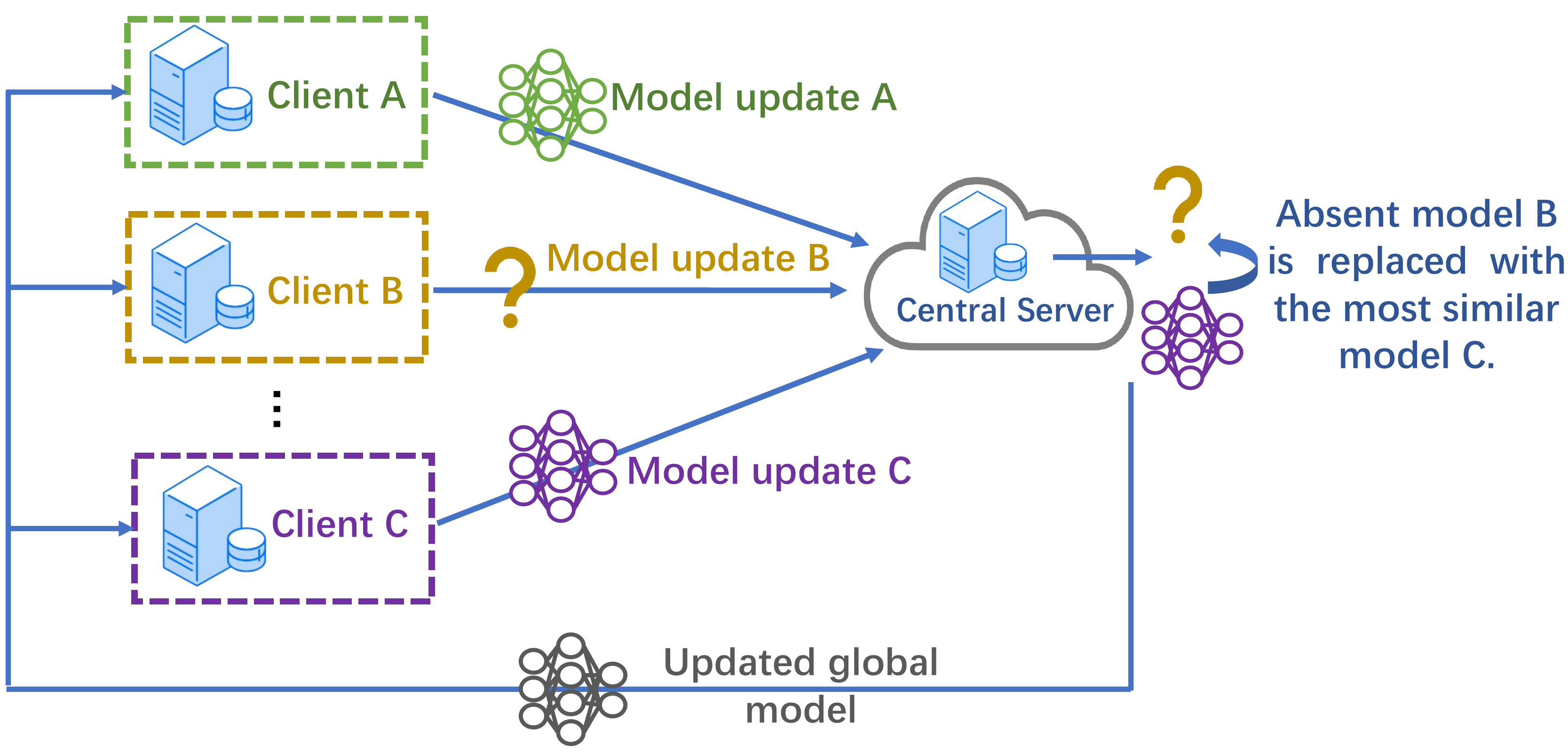}
    \vspace{-0.5cm}
    \caption{The schematic illustration of SAFARI.}
    \label{fedframe}
\end{figure}

In this section, we introduce our approach and elaborate on details of the proposed SAFARI algorithm as illustrated in~\cref{fedframe}.

\subsection{Core Concepts and Approach}
Here we first describe the two building blocks of the proposed SAFARI algorithm, which are \textit{the sparsity enabled communication efficiency} and \textit{the similarity assisted bias reduction with unreliable wireless communications}. The target problem and the proposed solution are explained in details.

\textbf{FL with Limited and Unreliable Communications:} According to the previous work, the lightweight message based connectionless protocol UDP is commonly used in resource-limited wireless communications. Specifically, UDP reduces much overhead by omitting mechanisms like ACK message confirmation and lost package retransmission~\cite{ye2022decentralized}. Therefore, despite the relatively low communication overhead, the transmission reliability can not be guaranteed in UDP transmissions. Assume a link reliability list $P = \{p_{1}^{t},\dots,p_{m}^{t} \}$, where $0\leq p_{i}^{t} \leq 1$ is the probability that the server successfully receives the local model $\boldsymbol{x}_{i, \tau}^{t}$ from client $i$ at the communication round $t$. In real-world scenarios, each server-client link's reliability could depend on several factors, i.e., the quality of the channel, the distance between the central server and the corresponding client, as well as the reliability of the client device. 

\textbf{Sparsity enabled Communication Efficiency and Similarity assisted Bias Reduction:} To save computing resources and training/inference time, sparse learning on large neural networks has been widely deployed in the deep learning field~\cite{lee2018snip,ma2021effective,mocanu2018scalable,you2020drawing,evci2020rigging}. When being introduced to FL scenarios, it can save the communication overload by reducing the amount of model weights to be sent. In this context, we propose to conduct the sparse learning at local clients, and utilize the similarity of sparse models to address the bias caused by unreliable communications. Concretely, the server keeps a record of the similarity across clients, which is measured by the sparse models they produce. The similarity record changes along with the training process according to the sparse models successfully received at each global round. With this record, for inactive client whose model has not been received by the server (client fails to participate in training or encounters network failure), the missing model is substituted by the model from the most similar active client. 

We will show in the theoretical part that in such way, the bias caused by random loss of local updates can be entirely eliminated when the clients are clusterable, or at least be limited to the same order of the intrinsic data dissimilarity bound under more general scenarios. This enables us to keep the same asymptotic convergence rate as vanilla FedAvg with perfectly reliable communications.

\subsection{The SAFARI Algorithm}
The proposed SAFARI algorithm to address the limited and unreliable communication issue is summarized in~\cref{alg:alg1}. As in vanilla FedAvg~\cite{mcmahan2017communication}, the server first initializes an original global model $\boldsymbol{x}^{0}$ and broadcasts it through communication links. Due to the unreliability of communications, some clients may fail to receive the global model from the server. For each client $i$ that successfully receives the global model, it first calculates a mask $\mathcal{M}_{i}$ based on a specific sparse algorithm to sparsify the global model's structure, and then performs  $\tau$ iterations' local SGD with the sparse structure. Once the local sparse training is completed, the client will send the sparse local model $\boldsymbol{x}^{t}_{i, \tau}$ back to the server, as illustrated in~\cref{alg:alg2}. 

\renewcommand{\algorithmicrequire}{\textbf{Input:}}
\renewcommand{\algorithmicensure}{\textbf{Initialize:}}
\begin{algorithm}[H]
\caption{SAFARI}\label{alg:alg1}
\begin{algorithmic}
    \STATE 
	\REQUIRE The number of communication rounds $T$, the learning rate $\eta $, the number of local steps $\tau$.
	\ENSURE The initial dense global model $\boldsymbol{x}^{0}$.
	\FOR {$t = 0$ to $T-1$}
	    \STATE Server broadcasts $\boldsymbol{x}^{t}$ to all clients.
	    \FOR {each client $i$ receives the message \textbf{in parallel}}
	    \STATE Perform \textit{Local Sparse Training}($\boldsymbol{x}^{t}, \eta, \tau$).
	    \STATE Send the updated sparse model $\boldsymbol{x}^{t}_{i, \tau}$ back to the server.
	    \ENDFOR
	    \STATE Server performs \textit{Bias Reduced Global Aggregation.} 
	\ENDFOR
	\STATE Finish the training with global model $\boldsymbol{x}^{T}$.
\end{algorithmic}
\end{algorithm}

\begin{algorithm}[H]
\caption{Local Sparse Training.}\label{alg:alg2}
\begin{algorithmic}
    \STATE 
	\REQUIRE The received global model $\boldsymbol{x}^{t}$, the learing rate $\eta$, the number of local steps $\tau$.
    \STATE Calculate mask $\mathcal{M}_{i}$ based on a specific sparse algorithm.
    \STATE Prune the model for a sparser structure: $\boldsymbol{x}^{t}_{i, 0}$ = $\boldsymbol{x}^{t} \odot \mathcal{M}_{i}$.
    \FOR {$k = 1$ to $\tau$}
        \STATE Sample a mini-batch $\xi_{i,k}$ from local dataset $D_i$.
        \STATE Compute the local gradient $g_{i}(\boldsymbol{x}^{t}_{i, 0} | \xi_{i,k})$.
        \STATE Local SGD step: $\boldsymbol{x}^{t}_{i, k} \!=\! \boldsymbol{x}^{t}_{i, k-1}\!-\!\frac{\eta}{\tau}g_{i}(\boldsymbol{x}^{t}_{i, k-1} | \xi_{i,k})$. 
    \ENDFOR
    \RETURN $\boldsymbol{x}_{i, \tau}^{t}$. 
\end{algorithmic}
\label{alg2}
\end{algorithm}

Again, since the communications are unreliable, not all of the updated local models can be received by the server. To address the potential bias caused by such random loss of client updates, the server will determine the active client group $\mathbb{M}_{a}$ based on the received client models. Before the aggregation, the server will update the similarity matrix among active clients, and then replace the model from each missing client $j$ with the received model from the most similar active client $j^\prime$, as shown in~\cref{alg:alg3}. After the total $T$ global rounds, the FL training is completed with a trained global model $\boldsymbol{x}^{T}$.

\begin{algorithm}[H]
\caption{Global Aggregation with Similarity-based Compensation.}\label{alg:alg3}
\begin{algorithmic}
    \STATE 
	\REQUIRE The received client models, the whole client set $\mathbb{M}$, the active client group $\mathbb{M}_{a}=\varnothing$, and $s$ similarity function. 
	\FOR {each client $i$ whose model has been received}
	\STATE $\mathbb{M}_{a}=\mathbb{M}_{a} \cup \left\{i \right\}$.
	\ENDFOR
	\STATE Server updates the similarity matrix $\rho \in \mathbb{R}^{m\times m}$ with $\rho_{u, v}=s(\boldsymbol{x}_{u, \tau}^{t}, \boldsymbol{x}_{v, \tau}^{t}), \forall u,v \in \mathbb{M}_{a}, u\ne v$.
	\FOR {each client $j \in \mathbb{M}\setminus\mathbb{M}_{a}$}
	\STATE $j' \leftarrow$ $i \in \mathbb{M}_{a}$ that maximizes $\rho_{i, j}$.
    \ENDFOR
    \STATE Server performs global aggregation: 
    \STATE $\boldsymbol{x}^{t+1}=\frac{1}{m} (\sum_{i\in \mathbb{M}_{a}}\boldsymbol{x}^{t}_{i, \tau} + \sum_{j\in \mathbb{M}\setminus\mathbb{M}_{a}} \boldsymbol{x}^{t}_{j', \tau}) $.
    \RETURN $\boldsymbol{x}^{t+1}$.
\end{algorithmic}
\label{alg3}
\end{algorithm}

\section{Theoretical Analysis}
In this section, we analyze the convergence of our method and
theoretically prove that it can achieve the same convergence rate as the vanilla FedAvg with reliable communications~\cite{wang2020tackling}.

\textbf{Notation.} In the following part, we use $\left \| \boldsymbol{x} \right \| $, $\left\| \boldsymbol{x} \right\|_1$ and $\left [ \boldsymbol{x} \right ] _n $ to denote the $l_2$, $l_1$ norms and the $n$-th element of a vector $\boldsymbol{x}$, respectively.

\subsection{Assumptions} 
\subsubsection{\textbf{Functions}} We first adopt the following three standard assumptions on functions, which are widely used in the non-convex federated optimization field:
\begin{itemize}
\item \textbf{Smoothness.} \textit{The local objective functions are L-smooth, i.e., $\forall i\in \mathbb{M}$}:
\begin{equation}
    \left\|\nabla \mathcal{L}_{i}(\boldsymbol{x})-\nabla \mathcal{L}_{i}(\boldsymbol{y})\right\| \leq L\|\boldsymbol{x}-\boldsymbol{y}\|, \forall \boldsymbol{x}, \boldsymbol{y} \in \mathbb{R}^{d}.
\end{equation}
\item \textbf{Unbiased Gradient and Bounded Variance.} \textit{$\forall i\in\mathbb{M}$, the stochastic gradient $g_i(\boldsymbol{x}|\xi)$ calculated with local data batch $\xi$ is an unbiased estimator of the local gradient: $\mathbb{E}_{\xi \sim D_{i}}\left[g_{i}(\boldsymbol{x} | \xi)\right]=\nabla \mathcal{L}_{i}(\boldsymbol{x})$, and the variance is bounded by: $\mathbb{E}_{\xi \sim D_{i}}\left\|g_{i}(\boldsymbol{x}|\xi)\!\!-\!\!\nabla \mathcal{L}_{i}(\boldsymbol{x})\right\|^{2} \leq \sigma^{2}$, $\forall \boldsymbol{x} \in \mathbb{R}^{d}$, $\sigma^{2} \geq 0$.}
\item \textbf{Bounded Dissimilarity.} \textit{There exist constants $\beta^2 \geq 1$ and $\zeta^2 \geq 0$ such that}:
\begin{equation}
    \frac{1}{m} \sum_{i=1}^{m} \left\|\nabla \mathcal{L}_{i}(\boldsymbol{x})\right\|^{2} \leq \beta^{2}\left\|\frac{1}{m} \sum_{i=1}^{m}\nabla \mathcal{L}_{i}(\boldsymbol{x})\right\|^{2}+\zeta^{2}.
\end{equation}
\textit{Particularly, $\beta^2 = 1$ and $\zeta^2 = 0$ indicate the IID situation where all the local functions are identical.}
\end{itemize} 

\subsubsection{\textbf{Sparse Models}} To analyze the property of local training with sparse models, a common assumption on the mask-induced error is also adopted from sparsification-related literature~\cite{ma2021effective}.

\begin{itemize}
    \item \textbf{Mask-induced Error.} \textit{It is assumed that $\forall \boldsymbol{x} \in \mathbb{R}^{d}$, the corresponding binary mask $\mathcal{M} \in \{0, 1\}^{d}$ satisfy
\begin{equation}
    \left\|\boldsymbol{x} \odot \mathcal{M}-\boldsymbol{x}\right\|^{2} \leq \delta^{2}\left\|\boldsymbol{x}\right\|^{2}, 0<\delta <1,
\end{equation}
where $\odot$ denotes the Hadamard product.
}
\end{itemize}

Note that the above assumption is a quite relaxed one, which is not limited to any specific sparse algorithms. Furthermore, to analyze the impact of sparse learning in distributed fashion, we make an assumption on the similarity between local training with sparse structures.   
\begin{itemize}
\item \textbf{Similarity Preservation.} \textit{Under the bounded dissimilarity assumption, $\forall \boldsymbol{x} \in \mathbb{R}^{d}$, $\forall i,j \in \mathbb{M}$ and local model mask $\left \{ \mathcal{M}_i \right \}_{i=1}^m$}:
\begin{equation}
 \left\|\nabla \mathcal{L}_{i}(\boldsymbol{x} \odot \mathcal{M}_i)\right\|^{2}\! \leq \! \beta^{2}\left\| \nabla \mathcal{L}_{j}(\boldsymbol{x} \odot \mathcal{M}_j)\right\|^{2}+\zeta^{2}.
\end{equation}
\end{itemize}

The above assumption indicates the rationality behind the compensation based on the similarity among sparse models produced by different clients. The theoretical analysis that demonstrates this assumption will hold for most existing sparse algorithms is provided in the appendix. 

\subsubsection{\textbf{Communication Networks}} Similar to previous work, we also make an additional assumption on the unreliable communication network~\cite{ye2022decentralized}. But compared with the \textit{independent and stable links} assumption made by Ye \textit{et al.}, we extend the condition to cover \textit{independent and unstable links}. In other words, the algorithm proposed in this paper does not require the link reliability to be known in advance or keep stable during training.
\begin{itemize}
    \item \textbf{Independent and Unstable Links.} \textit{The transmissions on different client links are independent and each link's reliability may change during training process.} 
\end{itemize}

\subsection{Descent Lemma with Sparsification}
\noindent \textbf{Lemma 1.} (Descent Lemma with Sparsification) \textit{With the above assumption on function smoothness, unbiased gradient and bounded variance, as well as sparsification, if $\eta \leq \tau/(6L)$, it holds $\forall i\in\mathbb{M}$, $t\in\mathbb{T}=\{0, \ldots, T-1\}$, $k\in\mathbb{K}$ that, 
\begin{align}
\mathbb{E}\left[\mathcal{L}_{i}(\boldsymbol{x}_{i, k}^{t})\right] & \!\leq\!  \mathbb{E}\left[\mathcal{L}_{i}(\boldsymbol{x}_{i, k\!-\!1}^{t})\right]\!-\!\frac{\eta}{3 \tau} \mathbb{E}\left\|\nabla \mathcal{L}_{i}(\boldsymbol{x}_{i, k\!-\!1}^{t})\right\|^{2} \nonumber \\
&+\frac{\eta^{2} L \sigma^{2}}{2 \tau^2}+\frac{2 \eta L^{2} \delta^{2}}{3\tau} \mathbb{E}\left\|\boldsymbol{x}_{i, k\!-\!1}^{t}\right\|^{2}.
\end{align}
}

From Lemma 1, with the appropriate learning rate, the local objective value will decrease by $\frac{\eta}{3\tau} \mathbb{E}\|\nabla \mathcal{L}_{i}(\boldsymbol{x}_{i, k-1}^{t})\|^{2}$ after every local step. The lemma also meets the expectation that the training will suffer from stochastic gradient variance $\sigma$ and weight pruning error $\delta$. To the best of the authors' knowledge, rigorous analysis to quantify such error is still uncertain in related research fields, and is also beyond the scope of this work. However, the empirical success of the popular sparse algorithms implies that the tolerance to such error can be quite large in practice~\cite{lee2018snip,ma2021effective,mocanu2018scalable,you2020drawing,evci2020rigging}, which enables us to implement sparse training in FL for communication efficiency, and meanwhile utilizes the properties of sparse models to address the unreliable communications. 

\subsection{Global Convergence}
To keep consistent and fair comparison with existing FL researches, we build our analysis within the general analysis framework for heterogeneous federated optimization algorithm proposed by~\cite{wang2020tackling}. The lemma below points out the influence of unreliable communication links and the following remark explains how our proposed method perfectly addresses such influence with similarity-maintaining sparse models in clusterable scenarios. \\
\noindent \textbf{Lemma 2.} \textit{Under the above assumptions, if $\eta \leq \frac{1}{2\tau L}$, then the optimization error will be bounded as follows}:
\begin{align}
&\frac{1}{T} \sum_{t=0}^{T-1} \mathbb{E}\left\|\nabla \mathcal{L}(\boldsymbol{x}^{t})\right\|^{2} 
\leq  \frac{4\left[\mathcal{L}(\boldsymbol{x}^{0})-\mathcal{L}_{\mathrm{inf}}\right]}{3\eta \tau T} \nonumber \\
& + \frac{16\tau \eta L\sigma^{2}}{3m} +2 \eta^{2} \sigma^{2} L^{2}(\tau-1)  +  4 \eta^{2} L^{2} \tau(\tau-1)\zeta^{2} \nonumber \\
&+(2 \tau \eta L-2/3) \frac{1}{m^{2}} \varphi ,
\end{align}
\textit{where $\varphi  = \sum_{i=1}^{m} \left(1-p_{i}^{t}\right)^{2}  \mathbb{E}\|\boldsymbol{h}_{i^{\prime}}^{(t)}-\boldsymbol{h}_{i}^{(t)}\|^{2}$. Specifically, $ \boldsymbol{h}_{i}^{(t)}=\frac{1}{\tau} \sum_{k=1}^{\tau}  \nabla \mathcal{L}_{i}(\boldsymbol{x}_{i, k}^{t})$, and $i'$ is the index of the most similar client used for replacing client $i$ in case it is lost. See the appendix for the definition and proof details.}

\noindent \textbf{Remark 1.} In Lemma 2, $\varphi $ captures the influence of the random communication network. If we keep the constraints on the learning rate unchanged as in the classic analysis~\cite{wang2020tackling}, then $\left(2 \tau \eta L-2/3\right) \geq 0$ is possible and the global convergence will suffer from $\varphi $ with any unreliable communication link, i.e., $p_i\leq 1$. However, with the proposed method, if the clients are clusterable, i.e., $\mathbb{E}\|\boldsymbol{h}_{i^{\prime}}^{(t)}-\boldsymbol{h}_{i}^{(t)}\|^{2}=0$, then $\varphi $ will be zero and thus the impact of unreliable communication links is entirely eliminated regardless of the unreliability. In this case, the convergence property of vanilla FedAvg with reliable communications will be perfectly preserved. It justifies our strategy to determine the alternative $\boldsymbol{h}_{i^{\prime}}^{(t)}$ based on sparse model similarity: the property that the drift between sparse models will be bounded by $\zeta$ enables us to reduce the variance caused by unreliable communication into that caused by non-IID data distribution, which is intrinsic in FL and can be addressed by a separate line of research works. See the appendix for proof details. 

Furthermore, the following theorem indicates the proposed method converge at the asymptotic rate as vanilla FedAvg with reliable communications even with unclusterable clients.

\noindent \textbf{Theorem 1} \textit{Under the above assumptions, if $\eta=\sqrt{\frac{m}{{\tau} T}}$, the optimization error after total $T$ iterations is bounded as follows}:
\begin{align}
    \min _{t \in\mathbb{T}} & \mathbb{E}\left\|\nabla \mathcal{L}(\boldsymbol{x}^{t})\right\|^{2} \!\leq\! \mathcal{O}\left(\frac{1}{\sqrt{m \tau T}}\right)\!+\!\mathcal{O}\left(\frac{A \sigma^{2}}{\sqrt{m \tau T}}\right) \nonumber \\
    &\!+\!\mathcal{O}\left(\frac{m B \sigma^{2}}{\tau T}\right)\!+\!\mathcal{O}\left(\frac{m C \zeta^{2}}{\tau T}\right),
\end{align}
\textit{where} $A = \tau, B = \tau-1,C = \tau(\tau-1)$ \textit{, and all other constants are subsumed in $\mathcal{O}$. Please refer to appendix for details.}


\textbf{Comparison with vanilla FedAvg.} Compared with the convergence analysis of FedAvg in~\cite{wang2020tackling}, the above theorem theoretically indicates that SAFARI with unreliable communications can achieve the same asymptotic convergence rate as FedAvg with reliable communication network under the same parameter setting. Hence, the negative influence of communication unreliability is effectively controlled. In the next section,  the experiment results that confirm our theoretical analysis are provided.

\section{Experiments}
We evaluate the proposed framework with different sparse algorithms with 10 clients. We train the ResNet-20 model~\cite{he2016deep} on the CIFAR-10 dataset, which contains 50,000 images for training and 10,000 images for testing. Specifically, the models are trained using Adam~\cite{kingma2015adam} optimizer with learning rate of 0.001, batch size of 64 and tested using batch size of 256. All of our experimental results are trained and evaluated using two NVIDIA-3090 GPUs with 24GB GPU RAM.


\subsection{Performance of SAFARI on Non-IID Data Distribution}
To evaluate the generalization of our framework, we have compared the performance of five representative neural network pruning algorithms with SAFARI. The sparsity level $\alpha$ is set to 80\%, where 80\% of model parameters will be pruned to 0. The selected pruning algorithms include: (1) Rand~\cite{tanaka2020pruning}: randomly prunes 80\% parameters; (2) MAG~\cite{tanaka2020pruning}: prunes the 80\% smallest absolute values of the model parameters; (3) GraSP~\cite{wang2019picking}: preserves 80\% of gradient flow through the network; (4) Synflow~\cite{tanaka2020pruning}: uses the synaptic saliency score to determine the importance of parameters in the network; (5) SNIP~\cite{lee2018snip}: refers to the discussion in Section II.

Following the balanced non-IID data partition setting~\cite{diao2020heterofl} in FL, 10 total clients are divided into 2 groups equally, and each client contains 5 labels in CIFAR-10. Besides, local steps $\tau = 5$ and local learning rate $\eta = 0.001$ are set to perform the local sparse training in~\cref{alg:alg2}.

In addition, as addressed in Section III, the successful transmission probability for each link $P$ are chosen as $\{1, 0.3, 0.3, 0.3, 0.3, 1, 0.3, 0.3, 0.3, 0.3\}$, where at least one client in each group will participate in every training process, while the other links has the failure probability of 0.3.

\begin{figure}[htbp]
    \centering

    \includegraphics[width = 0.85\linewidth]{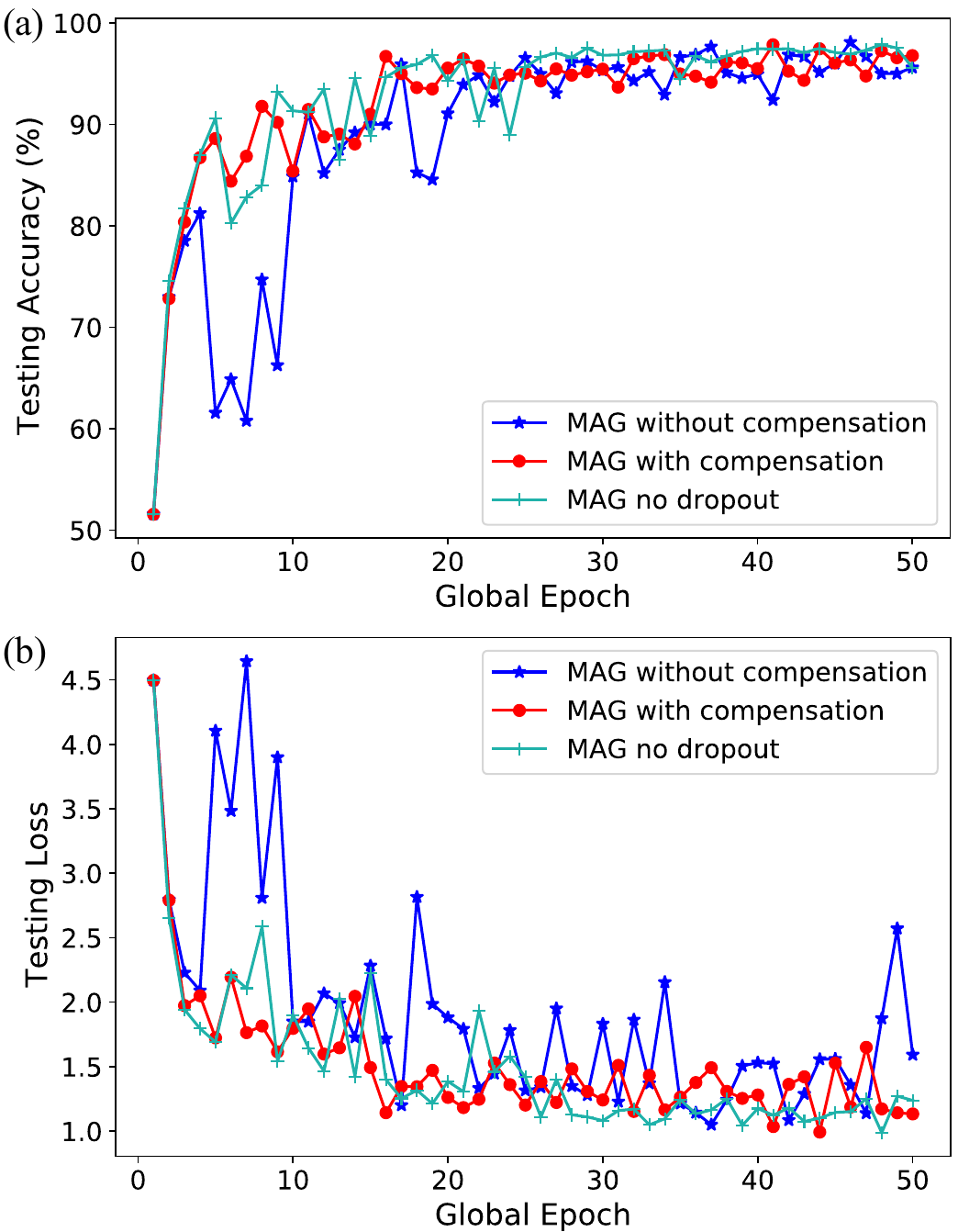}
    \vspace{-0.35 cm}
	\caption{Performance of SAFARI with MAG: (a) Testing accuracy; (b) Testing loss.} 
	\label{MAG_result}
\end{figure}

The results of MAG are shown in~\cref{MAG_result}. Without the proposed similarity-based compensation scheme for bias reduction, the unreliable communication channel will cause huge concussion during the global model training procedure. However, by introducing the compensation based on the similarity between sparse models, the model performance becomes more stable and accurate, which could reach 98\% (top-5) accuracy in CIFAR-10.

\begin{figure}[htbp]
    \centering
    \includegraphics[width = 0.85\linewidth]{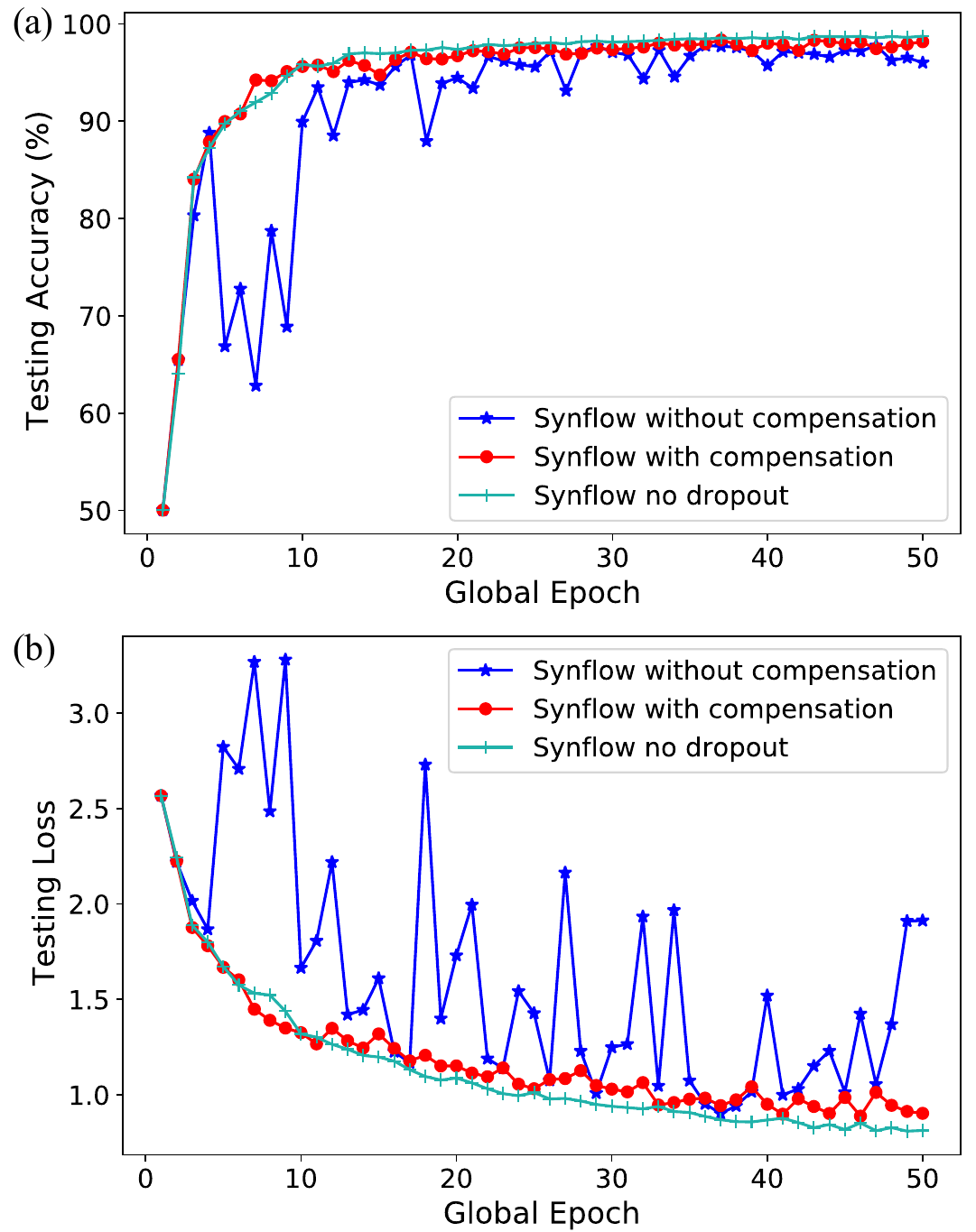}
    \vspace{-0.35 cm}
	\caption{Performance of SAFARI with Synflow: (a) Testing accuracy; (b) Testing loss.}
	
	\label{synflow_result}
\end{figure}

We also investigate the performance of SAFARI with Synflow. \cref{synflow_result} compares the convergence performance with respect to the number of iterations of Synflow training with and without compensation, as well as the original experiments with no dropouts. It can be seen that the training of Synflow with compensation has achieved nearly identical rate of convergence and convergence stability of Synflow with no dropouts, which is far superior to Synflow without compensation. The results of~\cref{synflow_result} exhibit consistent performance with~\cref{MAG_result}, which indicates the similarity-based compensation scheme will significantly improve the stability and speed of convergence.

\subsection{Validity of Similarity-based Compensation Scheme}
In this section, the experiments are conducted to verify the validity of the proposed similarity-based compensation scheme. Following the lemmas in Section IV, the $l_2$-norm based distance of model parameters of two clients $u, v$ is adopted in our experiment as the similarity function $s(\boldsymbol{x}_u,\boldsymbol{x}_v)$ in~\cref{alg:alg3}: 
\begin{equation}
s(\boldsymbol{x}_u,\boldsymbol{x}_v) := \|\boldsymbol{x}_u - \boldsymbol{x}_v\|.
\end{equation}

\begin{figure}[tbp]
    \centering
    \includegraphics[width = 0.75\linewidth]{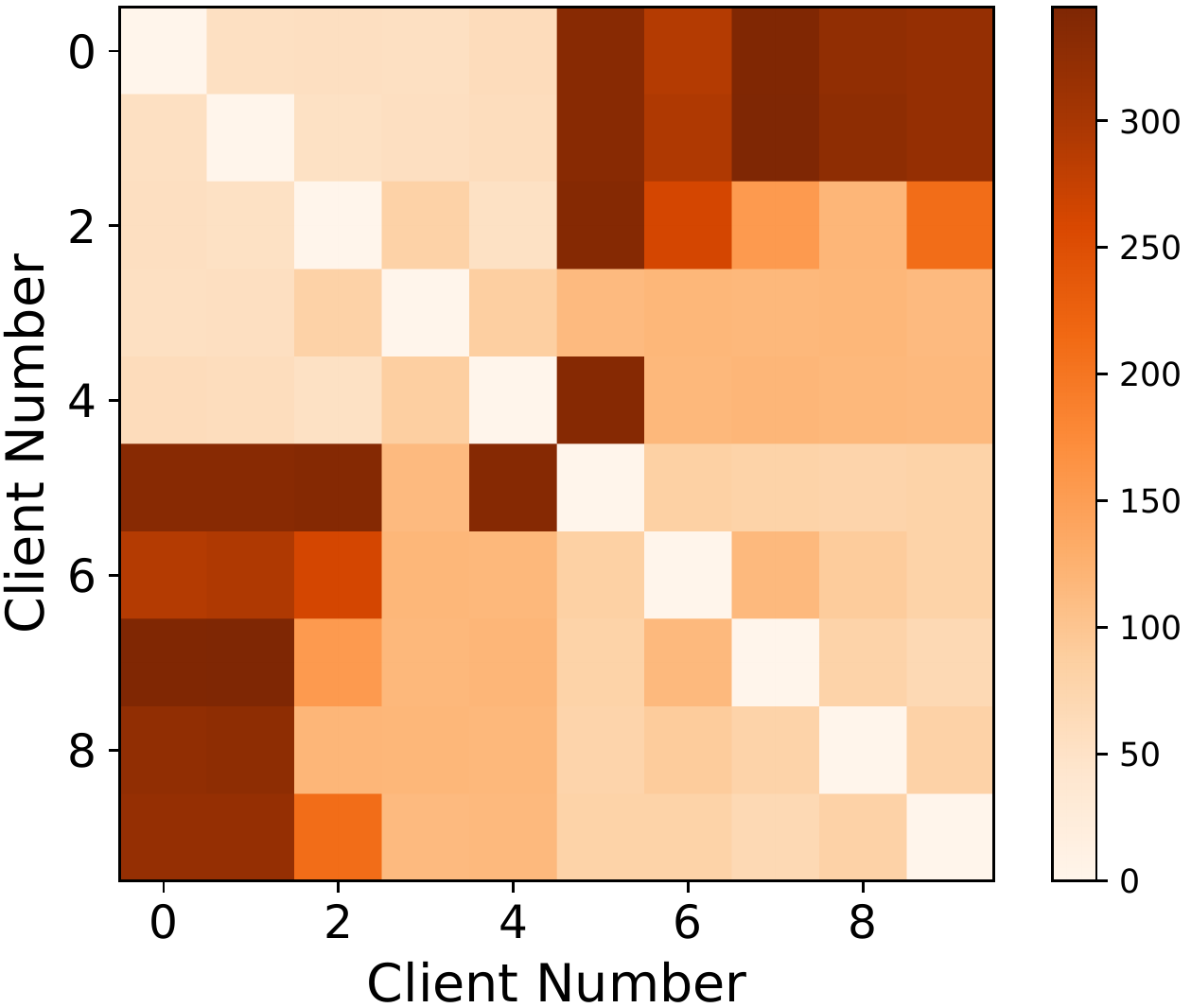}
    \vspace{-0.25  cm}
    \caption{The final similarity matrix.}
    \label{color_map}
\end{figure}

Particularly, we display the final similarity matrix $\rho$ after the whole training is completed, as plotted in~\cref{color_map}. For this experiment, following the basic setting, clients 0 to 4 are in Group 1 and have the same label split, while clients 5 to 9 are in Group 2. The lighter colored areas in the upper left and lower right corners indicate that the similarity between members of each group is relatively low. However, the areas in the lower left and upper right corners of this figure represent the similarity of clients between the two groups, and the dark colors between them indicate the high similarity. In the proposed scheme, since the client with the lightest colors in each line will be chosen to compensate, this figure proves that most of the time one client tends to select the model parameters in exactly the same group.

\begin{figure}[htbp]
    \centering
    \includegraphics[width = 0.85\linewidth]{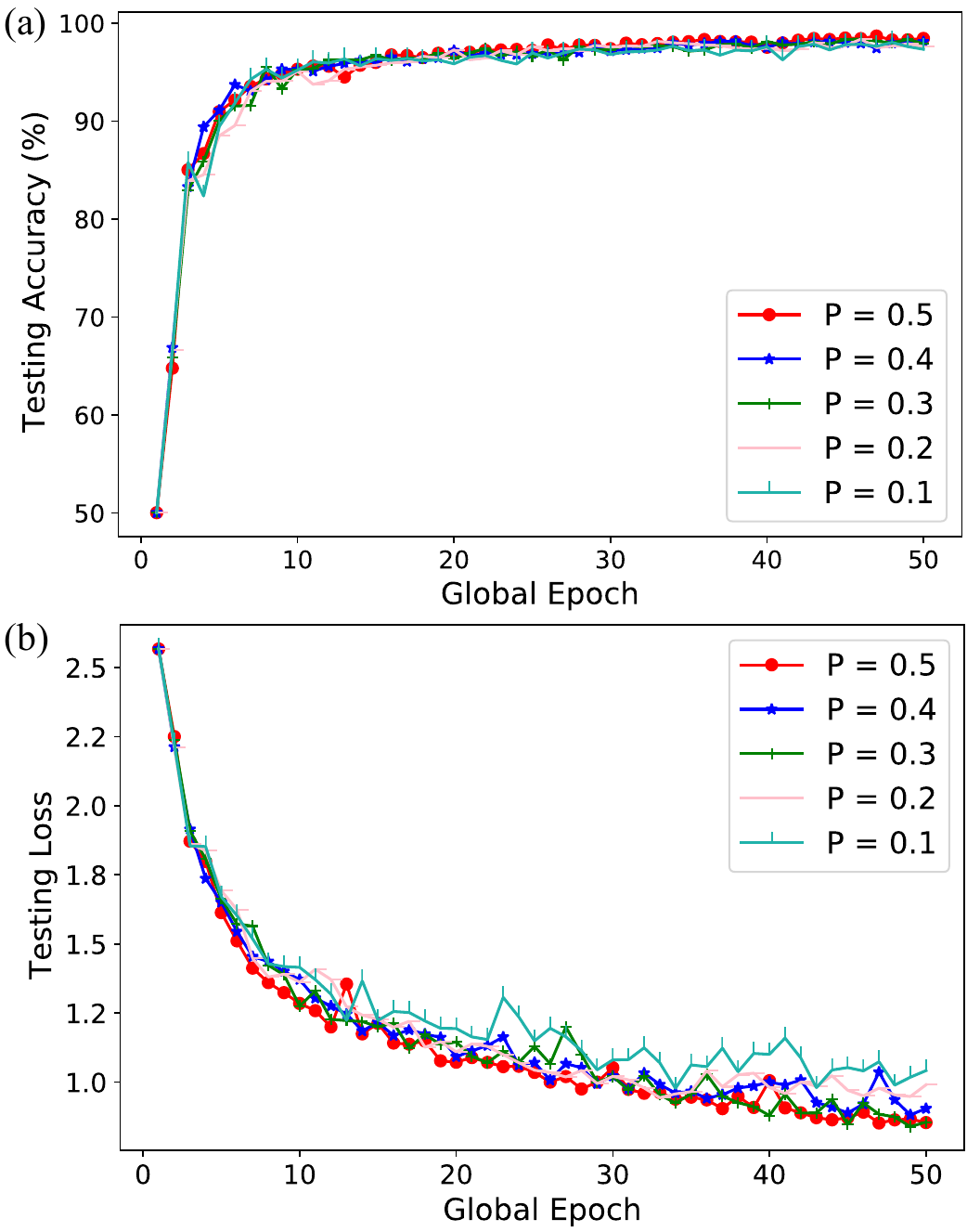}
    \vspace{-0.35 cm}
	\caption{Performance under different transmission probability settings: (a) Testing accuracy; (b) Testing loss. }
	\label{different_p}
\end{figure}

\begin{figure}[htbp]
    \centering
    \includegraphics[width = 0.85\linewidth]{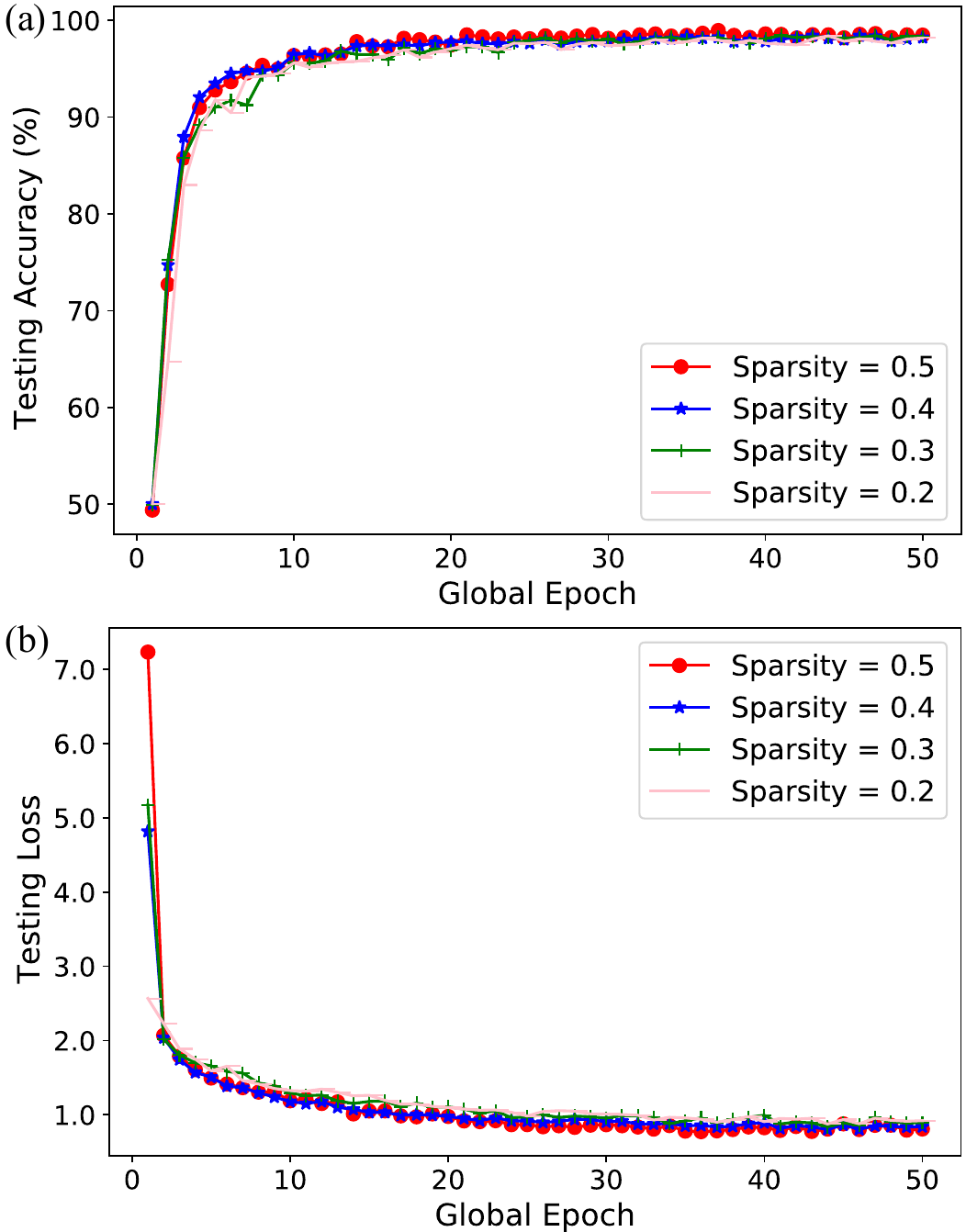}
    \vspace{-0.35 cm}
	\caption{Performance under different sparsity settings: (a) Testing accuracy; (b) Testing loss.  }
	\label{different_sparsity}
\end{figure}

\subsection{Evaluation of Stability}
In this section, we evaluate the validity and stability of SAFARI by changing two decisive hyperparameters: the successful transmission probability $P$ and the sparsity level $\alpha$. Specifically, only these two hyperparameters are changed and the other experiment settings are following the Synflow test. 

\subsubsection{Successful Transmission Probability}
In order to explore the performance of SAFARI under different $P$, five different $P$ are selected and the result is shown in~\cref{different_p}. We can observe that as the successful transmission probability $P$ changes, the testing accuracy will tend to be consistent, but the deduction in $P$ can significantly increase the volatility of testing loss.

\subsubsection{Sparsity Level} 
\cref{different_sparsity} shows the the training accuracy and training loss of SAFARI with different sparsity levels from 0.2 to 0.5. When the sparsity level $\alpha$ varies, We can find that the change in both the testing accuracy and testing loss is relatively small, and will eventually converge to a consistent interval.




\section{Conclusion}
In this paper, we propose a sparsity enabled robust FL framework, named as SAFARI, which can reduce communication overhead by local sparse learning, and meanwhile rectify the aggregation bias resulted from unreliable communications with unknown and potentially time-varying unreliability characteristics. Our theoretical analysis with respect to sparse model demonstrates that the similarity properties of client models are preserved under sparsity, and thus the proposed SAFARI algorithm with the similarity-based compensation can achieve the same asymptotic convergence rate as FedAvg with reliable communications. The experiments with CIFAR10 dataset and several representative sparse algorithms show that SAFARI can not only save up to 80$\%$ communication overhead but also consistently outperforms baselines by achieving fast and stable convergence under unreliable communications. Future work includes more sophisticated algorithm designs for more complex FL scenarios.



%

\appendix
In this section, we prove that under the bounded dissimilarity assumption, the local sparse models calculated by sparse learning methods maintain the relationship between local data distribution which is reflected by $\zeta$. 

Here we first analyze the masks calculated by sparse learning algorithms. Consider a client $i$ with data input $z_i$ sampled from local dataset $D_i$ and a dense neural network $\boldsymbol{x}\in \mathbb{R}^{d}$. Take the mask calculated with SNIP for example, the mask $\mathcal{M}_i\in \mathbb{R}^{d}$ for the given model $\boldsymbol{x}$ comes from the connection sensitivity of each model weight $[\boldsymbol{x}]_n, n = 1,\cdots,d$, which is defined as the effect of removing the connection,
\begin{align}
    [\Delta \mathcal{L}_{i}(\boldsymbol{x} ; D_i)]_{n} \!=\!\mathcal{L}_{i}(\mathbf{1} \odot \boldsymbol{x} ; D_i)-\mathcal{L}_{i}\left(\left(\mathbf{1}-\mathbf{e}_{n}\right) \odot \boldsymbol{x} ; D_i\right), \nonumber
\end{align}
where $\mathbf{e}_{n}$ is the one-hot indicating vector of the $n$-th element (i.e. zeros everywhere except at the index $n$) and $\mathbf{1}$ is a all-one vector with the same length of $\boldsymbol{x}$. 

To avoid the expensive $d+1$ forward passes over the dataset to calculate the precise $\Delta \mathcal{L}(\boldsymbol{x} ; D_i) \in \mathbb{R}^{d}$, the connection sensitivity is estimated by its infinitesimal version,
\begin{align}
    [\Delta \mathcal{L}_{i}(\boldsymbol{x} ; D_i)]_{n} \!\approx\! \lim _{\delta \rightarrow 0} \frac{\mathcal{L}_{i}(\mathbf{1}\! \odot\! \boldsymbol{x};D_i)\!\!-\!\!\mathcal{L}_{i}\left(\left(\mathbf{1}\!\!-\!\!\delta \mathbf{e}_{n}\right)\! \odot\! \boldsymbol{x} ; D_i\right)}{\delta}. \nonumber
\end{align}

Thus, the connection sensitivity is actually measured by the change
in loss with respect to an infinitesimal multiplicative perturbation $
\delta$ in weight $[\boldsymbol{x}]_n$. According to the SNIP algorithm design, this is computed by automatic differentiation in one
forward-backward pass~\cite{lee2018snip}. Note that the gradient with respect to the weight $[\boldsymbol{x}]_n$ is defined as numerical differentiation measured with respect to an additive change as follows,
\begin{align}
    [\nabla  \mathcal{L}_{i}(\boldsymbol{x};D_i)]_{n} \approx \lim _{\delta \rightarrow 0} \frac{\mathcal{L}_{i}( \boldsymbol{x}; D_i)-\mathcal{L}_{i}\left(\boldsymbol{x}+\delta \mathbf{e}_{n} ; D_i\right)}{\delta}.
\end{align}

Considering that automatic differentiation is usually used to avoid the error of numerical differentiation during back propagation, the connection sensitivity of the model weights can be regarded as being equivalent to the gradients calculated with the same dataset,
\begin{align}
    \Delta  \mathcal{L}_{i}(\boldsymbol{x} ; D_i) \approx \nabla  \mathcal{L}_{i}(\boldsymbol{x} ; D_i).
    \label{eq:relationship1}
\end{align}

Intuitively speaking, the weight with a higher connection sensitivity value has a considerable effect on the loss, and thus should be preserved. Therefore, the probability that the weight $[\boldsymbol{x}]_n$ is preserved after pruning ($[\mathcal{M}_i]_n = 1$) is is measured by the corresponding sensitivity magnitude,
\begin{align}
    s_{n}=\frac{\left|[\Delta  \mathcal{L}_{i}(\boldsymbol{x} ; D_i)]_n \right|}{\sum_{k=1}^{d}\left|[\Delta  \mathcal{L}_{i}(\boldsymbol{x} ; D_i)]_k\right|}, 
\end{align}
and there is,
\begin{align}
\mathbb{E}\left \| g_{i}\left(\boldsymbol{x}\! \odot \!\mathcal{M}_i  | z_{i}\right) \right \|_{1}  \!=\!\frac{\mathbb{E} \left \| g_{i}\left(\boldsymbol{x} | z_{i}\right)  \odot  \Delta  \mathcal{L}_{i}(\boldsymbol{x} ; D_i) \right \|_1}{\mathbb{E}\left \|\Delta  \mathcal{L}_{i}(\boldsymbol{x} ; D_i) \right \|_1}. \label{eq:relationship2}
\end{align}

The~\eqref{eq:relationship1} and~\eqref{eq:relationship2} indicate that the effect of mask depends only on the dense model weights and local data. It should be noted that in addition to SNIP, the masks from other sparse algorithms are also based only on data and dense models, except with much more direct algorithm designs. Next, assume the initialization of each model weight is independent, and we have
\begin{align}
    \left \| \nabla \mathcal{L}_{i}(\boldsymbol{x}\! \odot\! \mathcal{M}_i )\right \| _1 \! = \!\frac{\left \| \Delta  \mathcal{L}_{i}(\boldsymbol{x}) \right \|_1 \left \| \nabla \mathcal{L}_{i}(\boldsymbol{x}) \right \|_1 }{ \left \| \Delta  \mathcal{L}_{i}(\boldsymbol{x}) \right \|_1 } \!=\! \left \| \nabla \mathcal{L}_{i}(\boldsymbol{x}) \right \|_1 , \nonumber
\end{align}
where $\nabla  \mathcal{L}_{i}(\boldsymbol{x} ; D_i)$ and $\Delta  \mathcal{L}_{i}(\boldsymbol{x} ; D_i)$ are abbreviated as $\nabla  \mathcal{L}_{i}(\boldsymbol{x})$ and $\Delta  \mathcal{L}_{i}(\boldsymbol{x})$, respectively.

With the bounded dissimilarity assumption on the dense gradients, it holds that \begin{align}
    \frac{1}{m} \sum_{i=1}^{m} \left\|\nabla \mathcal{L}_{i}(\boldsymbol{x}\!\odot\! \mathcal{M}_{i})\right\|^{2}\! \leq\! \beta^{2}\left\|\frac{1}{m} \sum_{i=1}^{m}\nabla \mathcal{L}_{i}(\boldsymbol{x}\!\odot\! \mathcal{M}_{i})\right\|^{2}\!+\!\zeta^{2}. \nonumber
\end{align}




The above result implies that the gradients calculated with well-designed sparse structure maintains the relationship between the dense gradients. Specifically, within a client cluster with similar data distributions and the same global model $\boldsymbol{x}$, for any two clients, i.e. $i$ and $j$, the following conditions are satisfied,
\begin{align}
    \mathbb{E}_{z_i \sim D_{i}}\!\!\left\| g_{i}(\boldsymbol{x} \odot \mathcal{M}_{i} | z_i) \right\|  \! &= \! \mathbb{E}_{z_j \sim D_{j}}\!\!\left\| g_{j}(\boldsymbol{x} \odot \mathcal{M}_{j} | z_j) \right\| , \nonumber \\
        \left\| \nabla \mathcal{L}_{i}(\boldsymbol{x} \odot \mathcal{M}_{i})  \right\| &=   \left\| \nabla \mathcal{L}_{j}(\boldsymbol{x}\odot \mathcal{M}_{j})  \right\|, \nonumber
\end{align}
where $\mathcal{M}_{i}$ and $\mathcal{M}_{j}$ are the masks computed for clients $i$ and $j$. Obviously, if all the local functions are identical, the sparse gradients make the bounded dissimilarity assumption hold perfectly with $\beta=1$ and $\zeta = 0$,
\begin{equation}
     \frac{1}{m} \sum_{i=1}^{m} \left\| \nabla \mathcal{L}_{i}(\boldsymbol{x}\odot \mathcal{M}_{i})  \right\|^{2}\! =\! \left\|\frac{1}{m} \sum_{i=1}^{m} \nabla \mathcal{L}_{i}(\boldsymbol{x}\odot \mathcal{M}_{i})  \right\|^{2}.
\end{equation}



\subsection{Proof of Lemma 1}
Here we simplify the local gradient $g_{i}(\boldsymbol{x}_{i, k-1}^{t}|\xi_{i, k})$ calculated with batch $\xi_{i, k}$ from client $i$ as $g_{i}(\boldsymbol{x}_{i, k-1}^{t} )$. Since $\mathcal{L}_{i}$ is $L$-smooth (smoothness assumption), at the global iteration $t$ and local iteration $k$, it holds that for each client $i$,
\begin{align}
\mathcal{L}_{i} & (\boldsymbol{x}_{i, k}^{t}) \leq  \mathcal{L}_{i}(\boldsymbol{x}_{i, k-1}^{t}) + \frac{L}{2}\left\|\boldsymbol{x}_{i, k}^{t}-\boldsymbol{x}_{i, k-1}^{t}\right\|^{2} \nonumber \\
&+\left\langle\nabla \mathcal{L}_{i}(\boldsymbol{x}_{i, k-1}^{t}), \boldsymbol{x}_{i, k}^{t}-\boldsymbol{x}_{i, k-1}^{t}\right\rangle \\
=& \mathcal{L}_{i}(\boldsymbol{x}_{i, k-1}^{t})+\frac{\eta^{2} L}{2\tau^{2}}\left\|g_{i}(\boldsymbol{x}_{i, k-1}^{t}\odot \mathcal{M}_{i})\right\|^{2} \nonumber\\
&-\frac{\eta}{\tau}\left\langle\nabla \mathcal{L}_{i}(\boldsymbol{x}_{i, k-1}^{t}), g_{i}(\boldsymbol{x}_{i, k-1}^{t}\odot \mathcal{M}_{i}) \right\rangle. 
\label{eq:lemma1_1}
\end{align}
 
Besides, with the unbiased gradient and bounded variance assumption, it is obvious that
\begin{align}
    &\mathbb{E}\left\langle\nabla \mathcal{L}_{i}(\boldsymbol{x}_{i, k-1}^{t}), g_{i}(\boldsymbol{x}_{i, k-1}^{t}\odot \mathcal{M}_{i}) \right\rangle \nonumber\\
    &= \mathbb{E} \left\langle\nabla \mathcal{L}_{i}(\boldsymbol{x}_{i, k-1}^{t}), \nabla \mathcal{L}_{i}(\boldsymbol{x}_{i, k-1}^{t}\odot \mathcal{M}_{i}) \right\rangle,
    \label{eq:lemma1_2}
\end{align}

and
\begin{align}
    \mathbb{E}\|g_{i}(\boldsymbol{x}_{i, k-1}^{t}\! \odot\!  \mathcal{M}_{i})\|^{2} 
    \!\leq\!  \mathbb{E}\|\nabla \mathcal{L}_{i}(\boldsymbol{x}_{i, k-1}^{t} \!\odot\!  \mathcal{M}_{i})\|^{2}\! +\! \sigma^{2}.
    \label{eq:lemma1_3}
\end{align}

By taking the expectation over \eqref{eq:lemma1_1} and substituting \eqref{eq:lemma1_2} and \eqref{eq:lemma1_3} into \eqref{eq:lemma1_1}, we achieve
\begin{align}
\mathbb{E}&\left [ \mathcal{L}_{i}(\boldsymbol{x}_{i, k}^{t}) \right ]  
\leq \mathbb{E}\left [ \mathcal{L}_{i}(\boldsymbol{x}_{i, k-1}^{t}) \right ]\!  +\!\frac{\eta^{2} L \sigma^{2}}{2\tau^{2}}\! \nonumber\\
&+\!\frac{\eta^{2} L}{2\tau^{2}} \mathbb{E}\|\nabla \mathcal{L}_{i}(\boldsymbol{x}_{i, k-1}^{t}\odot \mathcal{M}_{i})\|^{2} \nonumber \\
& -\!\frac{\eta}{\tau}\mathbb{E}\|\nabla \mathcal{L}_{i}(\boldsymbol{x}_{i, k-1}^{t})\|^{2} \!\! -\! \! \frac{\eta}{\tau} \mathbb{E}\left\langle\nabla \mathcal{L}_{i}(\boldsymbol{x}_{i, k-1}^{t}), \mathcal{T}_{0}\right\rangle,
\label{eq:lemma1_4}
\end{align}
where $\mathcal{T}_{0} = \nabla \mathcal{L}_{i}(\boldsymbol{x}_{i, k-1}^{t} \odot \mathcal{M}_{i})-\nabla \mathcal{L}_{i}(\boldsymbol{x}_{i, k-1}^{t})$.

Note that,
\begin{align}
\mathbb{E}&\|\nabla \mathcal{L}_{i}(\boldsymbol{x}_{i, k-1}^{t} \odot \mathcal{M}_{i})\|^{2} \nonumber\\
\leq& 2 \mathbb{E}\|\nabla \mathcal{L}_{i}(\boldsymbol{x}_{i, k-1}^{t})\|^{2} \nonumber \\
&+2 \mathbb{E}\|\nabla \mathcal{L}_{i}(\boldsymbol{x}_{i, k-1}^{t})-\nabla \mathcal{L}_{i}(\boldsymbol{x}_{i, k-1}^{t} \odot \mathcal{M}_{i})\|^{2}\\
\leq& 2 \mathbb{E}\|\nabla \mathcal{L}_{i}(\boldsymbol{x}_{i, k-1}^{t})\|^{2} \nonumber \\
&+2 L^{2} \mathbb{E}\|\boldsymbol{x}_{i, k-1}^{t} \odot \mathcal{M}_{i}-\boldsymbol{x}_{i, k-1}^{t}\|^{2}\\
\leq& 2 \mathbb{E}\|\nabla \mathcal{L}_{i}(\boldsymbol{x}_{i, k-1}^{t})\|^{2}+2 L^{2} \delta^{2} \mathbb{E}\|\boldsymbol{x}_{i, k-1}^{t}\|^{2},
\label{eq:lemma1_5}
\end{align}
where the last two inequalities come from the smoothness assumption and the relaxed assumption on mask-induced error. Next, with $2\langle \boldsymbol{a}, \boldsymbol{b}\rangle \leq  \|\boldsymbol{a}\|^{2}+\|\boldsymbol{b}\|^{2}$, we have,
\begin{align}
-&\mathbb{E}\left\langle\nabla \mathcal{L}_{i}(\boldsymbol{x}_{i, k-1}^{t}), \mathcal{T}_{0} \right\rangle \leq \frac{1}{2} \mathbb{E}\|\nabla \mathcal{L}_{i}(\boldsymbol{x}_{i, k-1}^{t})\|^{2} \nonumber \\
&+\frac{1}{2} \mathbb{E}\|\nabla \mathcal{L}_{i}(\boldsymbol{x}_{i, k-1}^{t} \odot \mathcal{M}_{i})-\nabla \mathcal{L}_{i}(\boldsymbol{x}_{i, k-1}^{t})\|^{2} \\
\leq& \frac{1}{2} \mathbb{E}\|\nabla \mathcal{L}_{i}(\boldsymbol{x}_{i, k-1}^{t})\|^{2}+\frac{L^{2} \delta^{2}}{2} \mathbb{E}\|\boldsymbol{x}_{i, k-1}^{t}\|^{2}.
\label{eq:lemma1_6}
\end{align}

Substitute \eqref{eq:lemma1_5} and \eqref{eq:lemma1_6} into\eqref{eq:lemma1_4}, and by setting $\eta \leq \tau/(6L)$, we have
\begin{align}
\mathbb{E}& \left[\mathcal{L}_{i}(\boldsymbol{x}_{i, k}^{t}) \right] \leq  \mathbb{E}\left[\mathcal{L}_{i}(\boldsymbol{x}_{i, k-1}^{t})\right]+\frac{\eta^{2} L \sigma^{2}}{2\tau^2} \nonumber \\
&-\frac{\eta(\tau-2 \eta L)}{2\tau^2} \mathbb{E}\|\nabla \mathcal{L}_{i}(\boldsymbol{x}_{i, k-1}^{t})\|^{2} \nonumber \\
&+\frac{\eta L^{2} \delta^{2}(\tau+2 \eta L)}{2 \tau^2} \mathbb{E}\|\boldsymbol{x}_{i, k-1}^{t}\|^{2} \\
 \leq & \mathbb{E}\left[\mathcal{L}_{i}(\boldsymbol{x}_{i, k-1}^{t})\right]-\frac{\eta}{3 \tau} \mathbb{E}\|\nabla \mathcal{L}_{i}(\boldsymbol{x}_{i, k-1}^{t})\|^{2 } \nonumber\\
&+\frac{\eta^{2} L \sigma^{2}}{2 \tau^{2}}+\frac{2 \eta L^{2 } \delta^{2}}{3 \tau} \mathbb{E}\|\boldsymbol{x}_{i, k-1}^{t}\|^{2}.
\end{align}

The proof of Lemma 1 is completed.

\subsection{Proof of Lemma 2}
From the global point of view, $\boldsymbol{x}_{i, \tau}^{t}$ represents the local model sent to server after client $i$'s local iterations, which is supposed to be a sparse one. Recall that the global model is updated by the following rule under reliable communications:
\begin{equation}
    \boldsymbol{x}^{t+1}=\frac{1}{m}\sum_{i =1}^{m} \boldsymbol{x}_{i, \tau}^{t}= \boldsymbol{x}^{t}- \tau\eta \sum_{i=1}^{m}\frac{1}{m} \boldsymbol{d}_{i}^{(t)},
\end{equation}
where $\boldsymbol{d}_{i}^{(t)}=\frac{1}{\tau} \sum_{k=1}^{\tau}  g_{i}(\boldsymbol{x}_{i, k}^{t})$ is the normalized stochastic gradient at client $i$. Correspondingly, the normalized gradient at each client is defined as
\begin{align}
    \boldsymbol{h}_{i}^{(t)}=\frac{1}{\tau} \sum_{k=1}^{\tau}  \nabla \mathcal{L}_{i}(\boldsymbol{x}_{i, k}^{t}) , i \in \mathbb{M}.
    \label{eq:h_i_t}
\end{align}

To solve the problem caused by unreliable communications, the global model is updated with the proposed compensation based on sparse model similarity. Therefore, the expectation of global model update can be written as 
\begin{equation}
    \mathbb{E} \left [ \boldsymbol{x}^{t+1}\!-\!\boldsymbol{x}^{t} \right ] \!=\! -\tau \eta \sum_{i=1}^{m} \frac{1}{m}\left [ p_{i}^{t} \boldsymbol{d}_{i}^{(t)}\!+\!\left ( 1\!-\!p_{i}^{t} \right )\boldsymbol{d}_{i'}^{(t)} \right ], 
\end{equation}
where $i'$ is the index of the most similar client used for replacing client $i$ in case it is lost, and $p_i^t$ is the reliability of the channel between client $i$ and the server at round $t$. 

According to the smoothness assumption, there is,
\begin{align}
\mathbb{E}&\left[\mathcal{L}(\boldsymbol{x}^{t+1})\right]\!-\!\mathcal{L}(\boldsymbol{x}^{t}) \nonumber \\
 \leq& \mathbb{E}\left\langle\nabla \mathcal{L}(\boldsymbol{x}^{t}), \boldsymbol{x}^{t+1}\!-\!\boldsymbol{x}^{t}\right\rangle  \!+\!\frac{L}{2} \mathbb{E}\|\boldsymbol{x}^{t+1}\!-\!\boldsymbol{x}^{t}\|^{2}\\
\leq&-\tau \eta \underbrace{\mathbb{E}\left\langle\nabla \mathcal{L}(\boldsymbol{x}^{t}), \sum_{i=1}^{m} \frac{1}{m}\left [ p_{i}^{t} \boldsymbol{d}_{i}^{(t)}\!+\!\left ( 1\!-\!p_{i}^{t} \right )\boldsymbol{d}_{i'}^{(t)} \right ]  \right\rangle}_{T_{1}}  \nonumber \\
&\!+\!\frac{\tau^{2} \eta^{2} L}{2} \underbrace{\mathbb{E}\left\|\sum_{i=1}^{m} \frac{1}{m}\left [ p_{i}^{t} \boldsymbol{d}_{i}^{(t)}\!+\!\left ( 1\!-\!p_{i}^{t} \right )\boldsymbol{d}_{i'}^{(t)} \right ] \right\|^{2}}_{T_{2}}.
\label{eq:lemma2_1}
\end{align}

\textbf{Bounding the first term $T_1$}. For the first term on the right hand side of~\eqref{eq:lemma2_1}, there is
\begin{align}
T_{1}
=&\mathbb{E}\left\langle\nabla \mathcal{L}(\boldsymbol{x}^{t}), \sum_{i=1}^{m}\frac{1}{m} p_{i}^{t}(\boldsymbol{d}_{i}^{(t)}\!-\!\boldsymbol{h}_{i}^{(t)}\!+\!\boldsymbol{h}_{i}^{(t)})\right\rangle \nonumber \\
&+\mathbb{E}\left\langle\nabla \mathcal{L}(\boldsymbol{x}^{t}), \sum_{i=1}^{m}\frac{1}{m} \left(1\!-\!p_{i}^{t}\right)(\boldsymbol{d}_{i'}^{(t)}\!-\!\boldsymbol{h}_{i'}^{(t)}+\boldsymbol{h}_{i'}^{(t)})\right\rangle \nonumber \\ 
=&\mathbb{E}\left\langle\nabla \mathcal{L}(\boldsymbol{x}^{t}), \sum_{i=1}^{m}\frac{1}{m} p_{i}^{t} \boldsymbol{h}_{i}^{(t)}\right\rangle \nonumber \\
&+ \mathbb{E}\left\langle\nabla \mathcal{L}(\boldsymbol{x}^{t}), \sum_{i=1}^{m}\frac{1}{m} \left(1\!-\!p_{i}^{t}\right) \boldsymbol{h}_{i'}^{(t)}\right\rangle,
\end{align}
where the second equality comes from the unbiased gradient assumption which implies $\mathbb{E} (\boldsymbol{d}_{i}^{(t)}\!-\!\boldsymbol{h}_{i}^{(t)} ) =0$.

\begin{align}
T_{1} =&\mathbb{E}\left\langle\nabla \mathcal{L}(\boldsymbol{x}^{t}), \sum_{i=1}^{m}\frac{1}{m} p_{i}^{t} \boldsymbol{h}_{i}^{(t)}\!+\!\sum_{i=1}^{m}\frac{1}{m} \left(1\!-\!p_{i}^{t}\right) \boldsymbol{h}_{i}^{(t)} \right\rangle \nonumber \\
&\!+\!\mathbb{E}\left\langle\nabla \mathcal{L}(\boldsymbol{x}^{t}), \sum_{i=1}^{m}\frac{1}{m} \left(1\!-\!p_{i}^{t}\right)(\boldsymbol{h}_{i'}^{(t)}\!-\!\boldsymbol{h}_{i}^{(t)}) \right\rangle \nonumber \\
=&\mathbb{E}\left\langle\nabla \mathcal{L}(\boldsymbol{x}^{t}), \sum_{i=1}^{m}\frac{1}{m} \boldsymbol{h}_{i}^{(t)}\right\rangle\nonumber \\
&+ \mathbb{E}\left\langle\nabla \mathcal{L}(\boldsymbol{x}^{t}), \sum_{i=1}^{m}\frac{1}{m} \left(1\!-\!p_{i}^{t}\right)(\boldsymbol{h}_{i'}^{(t)}\!-\!\boldsymbol{h}_{i}^{(t)})\right\rangle \nonumber \\
=&\frac{1}{2}\left\|\nabla \mathcal{L}(\boldsymbol{x}^{t})\right\|^{2}\!+\!\frac{1}{2} \mathbb{E}\left\|\sum_{i=1}^{m}\frac{1}{m} \boldsymbol{h}_{i}^{(t)}\right\|^{2} \nonumber \\
& +\mathbb{E}\left\langle\nabla \mathcal{L}(\boldsymbol{x}^{t}), \sum_{i=1}^{m}\frac{1}{m} \left(1\!-\!p_{i}^{t}\right)(\boldsymbol{h}_{i'}^{(t)}\!-\!\boldsymbol{h}_{i}^{(t)})\right\rangle \nonumber \\
&-\frac{1}{2} \mathbb{E}\left\|\nabla \mathcal{L}(\boldsymbol{x}^{t})\!-\!\sum_{i=1}^{m} \frac{1}{m}  \boldsymbol{h}_{i}^{(t)}\right\|^{2} \\
\leq& \frac{1}{2}\left\|\nabla \mathcal{L}(\boldsymbol{x}^{t})\right\|^{2}+\frac{1}{2} \mathbb{E}\left\|\sum_{i=1}^{m} \frac{1}{m} \boldsymbol{h}_{i}^{(t)}\right\|^{2} \nonumber \\
&\! +\! \frac{1}{2}\left\|\nabla \mathcal{L}(\boldsymbol{x}^{t})\right\|^{2} \!\! \!+ \! \! \frac{1}{2} \mathbb{E}\left\|\sum_{i=1}^{m}\!\frac{1}{m} \! \left(1\!-\!p_{i}^{t}\right)(\boldsymbol{h}_{i'}^{(t)}\!-\!\boldsymbol{h}_{i}^{(t)})\right\|^{2} \nonumber \\
&-\frac{1}{2} \mathbb{E}\left\|\nabla \mathcal{L}(\boldsymbol{x}^{t})-\sum_{i=1}^{m} \frac{1}{m} \boldsymbol{h}_{i}^{(t)}\right\|^{2}, 
\label{eq:T1}
\end{align}
where the last inequality follows from $2\langle \boldsymbol{a}, \boldsymbol{b}\rangle=\|\boldsymbol{a}\|^{2}+\|\boldsymbol{b}\|^{2}-\|\boldsymbol{a}-\boldsymbol{b}\|^{2}$.

\textbf{Bounding the second term $T_2$}. For the second term on the right hand side of\eqref{eq:lemma2_1}, there is,
\begin{align}
T_{2} 
=&\mathbb{E}\left\|\sum_{i=1}^{m} \frac{1}{m}  \boldsymbol{d}_{i}^{(t)} \right\|^{2} \nonumber \!\!\!+\! \mathbb{E}\left\|\sum_{i=1}^{m} \frac{1}{m} \left [ \left ( 1\!-\!p_{i}^{t} \right )(\boldsymbol{d}_{i'}^{(t)}\!-\!\boldsymbol{d}_{i}^{(t)} )\right ] \right\|^{2} \nonumber\\
\leq&  2 \mathbb{E}\left\|\sum_{i=1}^{m} \frac{1}{m} \boldsymbol{h}_{i}^{(t)}\right\|^{2} + 2 \mathbb{E}\left\|\sum_{i=1}^{m} \frac{1}{m} (\boldsymbol{d}_{i}^{(t)}-\boldsymbol{h}_{i}^{(t)})\right\|^{2} \nonumber \\
& + 3 \mathbb{E}\left\|\sum_{i=1}^{m} \frac{1}{m}\left ( 1-p_{i}^{t} \right )(\boldsymbol{h}_{i'}^{(t)}-\boldsymbol{h}_{i}^{(t)})\right\|^{2} \nonumber\\
&+3 \mathbb{E}\left\|\sum_{i=1}^{m} \frac{1}{m} \left ( 1-p_{i}^{t} \right )(\boldsymbol{h}_{i}^{(t)}-\boldsymbol{d}_{i}^{(t)})\right\|^{2} \nonumber \\
&+3 \mathbb{E}\left\|\sum_{i=1}^{m} \frac{1}{m} \left ( 1-p_{i}^{t} \right )(\boldsymbol{d}_{i'}^{(t)}-\boldsymbol{h}_{i'}^{(t)})\right\|^{2} \\
=& 2 \mathbb{E}\left\|\sum_{i=1}^{m}  \frac{1}{m} \boldsymbol{h}_{i}^{(t)}\right\|^{2} + 2 \sum_{i=1}^{m} \frac{1}{m^2} \mathbb{E}\|\boldsymbol{d}_{i}^{(t)}-\boldsymbol{h}_{i}^{(t)}\|^{2} \nonumber \\
&+3 \sum_{i=1}^{m}\frac{1}{m^2} \left ( 1-p_{i}^{t} \right )^{2} \mathbb{E}\|\boldsymbol{h}_{i'}^{(t)}-\boldsymbol{h}_{i}^{(t)}\|^{2} \nonumber \\
&+3 \sum_{i=1}^{m}\frac{1}{m^2} \left ( 1-p_{i}^{t} \right )^{2}\mathbb{E}\|\boldsymbol{h}_{i}^{(t)}-\boldsymbol{d}_{i}^{(t)}\|^{2} \nonumber \\
&+3\sum_{i=1}^{m}\frac{1}{m^2} \left ( 1-p_{i}^{t} \right )^{2} \mathbb{E}\|\boldsymbol{d}_{i'}^{(t)}-\boldsymbol{h}_{i'}^{(t)}\|^{2},
\end{align}
where the second equality is based on the independence of the surrogate selection, and the following inequalities follow from:$\left \| \sum_{i=1}^{n}\boldsymbol{w}_{i} \right \|_{2}^{2}\leq n\sum_{i=1}^{n}\left \| \boldsymbol{w}_{i} \right \|_{2}^{2}$.

With assumption on gradient variance, the second term can be bounded as,
\begin{align}
    T_{2} \leq& \sum_{i=1}^{m} \frac{1}{m^2} \left[ 2+6\left ( 1-p_{i}^{t} \right )^{2} \right]   \sigma ^{2} + 2 \mathbb{E}\left\|\sum_{i=1}^{m}  \frac{1}{m} \boldsymbol{h}_{i}^{(t)}\right\|^{2} \nonumber \\
    & + 3 \sum_{i=1}^{m}\frac{1}{m^2} \left ( 1-p_{i}^{t} \right )^{2} \mathbb{E}\|\boldsymbol{h}_{i}^{(t)}-\boldsymbol{h}_{i'}^{(t)}\|^{2}.
    \label{eq:T2}
\end{align}

\textbf{Bounding the objective reduction \eqref{eq:lemma2_1}.} Plugging \eqref{eq:T1} and \eqref{eq:T2} back into \eqref{eq:lemma2_1}, there is,
\begin{align}
\mathbb{E}&\left[\mathcal{L}(\boldsymbol{x}^{t+1})\right]-\mathcal{L}(\boldsymbol{x}^{t}) \nonumber \\
\leq& -\tau \eta \left\|\nabla \mathcal{L}(\boldsymbol{x}^{t})\right\|^{2}+\left ( \tau^{2} \eta^{2} L \!-\!\frac{\tau \eta}{2} \right ) \mathbb{E}\left\|\sum_{i=1}^{m}\frac{1}{m}  \boldsymbol{h}_{i}^{(t)}\right\|^{2} \nonumber \\
&+\frac{\tau \eta}{2}  \mathbb{E}\left\|\nabla \mathcal{L}(\boldsymbol{x}^{t})\!-\!\sum_{i=1}^{m} \frac{1}{m} \boldsymbol{h}_{i}^{(t)}\right\|^{2} \nonumber \\
& + \left (\frac{3}{2} \tau^{2} \eta^{2} L  \!-\!\frac{\tau \eta}{2} \right ) \sum_{i=1}^{m}\frac{1}{m^2} \left(1\!-\!p_{i}^{t}\right)^{2}  \mathbb{E}\|\boldsymbol{h}_{i'}^{(t)}\!-\!\boldsymbol{h}_{i}^{(t)}\|^{2} \nonumber \\
& + \tau^{2} \eta^{2} L \sum_{i=1}^{m} \frac{1}{m^2} \left[ 1+3\left ( 1-p_{i}^{t} \right )^{2} \right]   \sigma ^{2}.
\end{align}

If $\tau\eta L\leq \frac{1}{2}$, we have,
\begin{align}
    &\frac{1}{\tau\eta} \left( \mathbb{E}\left[\mathcal{L}(\boldsymbol{x}^{t+1})\right]-\mathcal{L}(\boldsymbol{x}^{t})\right)  \nonumber\\
    &\leq -\left\|\nabla \mathcal{L}(\boldsymbol{x}^{t})\right\|^{2} + \tau \eta L \sum_{i=1}^{m} \frac{1}{m^2} \left [ 1+3\left ( 1-p_{i}^{t} \right )^{2} \right ]   \sigma ^{2} \nonumber \\
    & + \left (\frac{3}{2} \tau \eta L   -\frac{1}{2} \right ) \sum_{i=1}^{m}\frac{1}{m^2} \left(1-p_{i}^{t}\right)^{2}  \mathbb{E}\|\boldsymbol{h}_{i'}^{(t)}-\boldsymbol{h}_{i}^{(t)}\|^{2} \nonumber \\
    &+\frac{1}{2}  \mathbb{E}\left\|\nabla \mathcal{L}(\boldsymbol{x}^{t})-\sum_{i=1}^{m} \frac{1}{m} \boldsymbol{h}_{i}^{(t)}\right\|^{2}  \\
    & \leq -\left\|\nabla \mathcal{L}(\boldsymbol{x}^{t})\right\|^{2} + \tau \eta L \sum_{i=1}^{m} \frac{1}{m^2} \left [ 1+3\left ( 1-p_{i}^{t} \right )^{2} \right ]   \sigma ^{2} \nonumber \\
    & + \left (\frac{3}{2} \tau \eta L   -\frac{1}{2} \right ) \sum_{i=1}^{m}\frac{1}{m^2} \left(1-p_{i}^{t}\right)^{2}  \mathbb{E}\|\boldsymbol{h}_{i'}^{(t)}-\boldsymbol{h}_{i}^{(t)}\|^{2} \nonumber \\
    & +\frac{1}{2}  \sum_{i=1}^{m} \frac{1}{m} \mathbb{E}\|\nabla \mathcal{L}_{i}(\boldsymbol{x}^{t})-\boldsymbol{h}_{i}^{(t)}\|^{2},
    \label{eq:bound on learning progress_1}
\end{align}
where the last inequality comes from Jensen’s Inequality $\left\|\sum_{i=1}^{m} \frac{1}{m}  \boldsymbol{w}_{i}\right\|^{2} \leq \sum_{i=1}^{m} \frac{1}{m}\left\|\boldsymbol{w}_{i}\right\|^{2} $.

\textbf{Bounding the difference between global gradient and normalized client gradient.} According to the definition of $\boldsymbol{h}_{i}^{(t)}$ in~\eqref{eq:h_i_t} and the smoothness assumption, there is
\begin{align}
&\mathbb{E}\|\nabla \mathcal{L}_{i}(\boldsymbol{x}^{t})-\boldsymbol{h}_{i}^{(t)}\|^{2} \nonumber \\
&= \mathbb{E}\|\frac{1}{\tau} \sum_{k=1}^{\tau}\left[\nabla \mathcal{L}_{i}(\boldsymbol{x}^{t})-\nabla \mathcal{L}_{i}(\boldsymbol{x}_{i, k}^{t})\right]\|^{2} \\
& \leq \frac{1}{\tau} \sum_{k=1}^{\tau} \mathbb{E}\left\|\nabla \mathcal{L}_{i}(\boldsymbol{x}^{t})-\nabla \mathcal{L}_{i}(\boldsymbol{x}_{i, k}^{t})\right\|^{2} \\
& \leq \frac{L^{2}}{\tau} \sum_{k=1}^{\tau} \mathbb{E}\|\boldsymbol{x}^{t}-\boldsymbol{x}_{i, k}^{t}\|^{2}.
\end{align}

In terms of bounding the difference between the global model $\boldsymbol{x}^{t}$ and the local model $\boldsymbol{x}_{i, k}^{t}$ after the $k$-th local iteration, we have the following results based on the model update rule as well as the assumption on the mask-induced error,
\begin{align}
\mathbb{E}&\|\boldsymbol{x}^{t}-\boldsymbol{x}_{i, k}^{t}\|^{2} = \eta^{2} \mathbb{E}  {\left\|\sum_{s=0}^{k-1} g_{i}(\boldsymbol{x}_{i, s}^{t}\odot \mathcal{M}_{i})\right\|^{2} } \nonumber \\
\leq& 2 \eta^{2} \mathbb{E} \left\|\sum_{s=0}^{k-1} \left[g_{i}(\boldsymbol{x}_{i, s}^{t}\odot \mathcal{M}_{i})-\nabla \mathcal{L}_{i}(\boldsymbol{x}_{i, s}^{t}\odot \mathcal{M}_{i})\right] \right\|^{2}  \nonumber \\
&+2 \eta^{2} \mathbb{E} {\left\| \sum_{s=0}^{k-1}  \nabla \mathcal{L}_{i}(\boldsymbol{x}_{i, s}^{t}\odot \mathcal{M}_{i})\right\|^{2} } \\
\leq& 2 \eta^{2} \delta^2 \mathbb{E} {\left\| \sum_{s=0}^{k-1}  \nabla \mathcal{L}_{i}(\boldsymbol{x, s}_{i}^{t})\right\|^{2} } + 2 \eta^{2} \sigma ^{2} k  \\
\leq& 2 \eta^{2}k \sum_{s=0}^{\tau-1} \mathbb{E}  {\|  \nabla \mathcal{L}_{i}(\boldsymbol{x}_{i, s}^{t})\|^{2} } + 2 \eta^{2} \sigma ^{2} k.
\end{align}

Taking the average over $\tau$ local iterations, we get,
\begin{align}
&\frac{1}{\tau} \sum_{k=1}^{\tau} \mathbb{E} {\|\boldsymbol{x}^{t}-\boldsymbol{x}_{i, k}^{t}\|^{2}}  \\
& \leq 2 \eta^{2} \sigma^{2}\left( \tau-1 \right) +2 \eta^{2}\left( \tau-1 \right) \sum_{k=0}^{\tau-1} \mathbb{E}\|\nabla \mathcal{L}_{i}(\boldsymbol{x}_{i, k}^{t})\|^{2}.
\label{eq:bound on global and local models}
\end{align}

Moreover, the local normalized gradient can be bounded by,
\begin{align}
 \mathbb{E}&\|\nabla \mathcal{L}_{i}(\boldsymbol{x}_{i, k}^{t})\|^{2}  \nonumber \\
 \leq& 2 \mathbb{E}\|\nabla \mathcal{L}_{i}(\boldsymbol{x}_{i, k}^{t})-\nabla \mathcal{L}_{i}(\boldsymbol{x}^{t})\|^{2} \nonumber +2 \mathbb{E}\|\nabla \mathcal{L}_{i}(\boldsymbol{x}^{t})\|^{2} \nonumber \\
 \leq& 2 L^{2} \mathbb{E}\|\boldsymbol{x}^{t}-\boldsymbol{x}_{i, k}^{t}\|^{2}  + 2 \mathbb{E}\|\nabla \mathcal{L}_{i}(\boldsymbol{x}^{t})\|^{2}.
\label{eq:bound on normalized gradient}
\end{align}

Plug \eqref{eq:bound on normalized gradient} back into \eqref{eq:bound on global and local models}, we achieve,
\begin{align}
\frac{1}{\tau} \sum_{k=1}^{\tau} \mathbb{E} & {\|\boldsymbol{x}^{t}-\boldsymbol{x}_{i, k}^{t}\|^{2} \leq 2 \eta^{2} \sigma^{2}\left( \tau-1 \right) } \nonumber \\
&+4 \eta^{2}L^{2}\left( \tau-1 \right) \sum_{k=0}^{\tau-1}  \mathbb{E}\|\boldsymbol{x}^{t}-\boldsymbol{x}_{i, k}^{t}\|^{2} \nonumber \\
& +4 \eta^{2}\left( \tau-1 \right) \sum_{k=0}^{\tau-1} \mathbb{E}\|\nabla \mathcal{L}_{i}(\boldsymbol{x}^{t})\|^{2}.
\end{align}

After rearranging, the difference between the global gradient and client gradient can be bounded by
\begin{align}
\frac{1}{\tau } \sum_{k=1}^{\tau} &\mathbb{E}\|\boldsymbol{x}^{t}-\boldsymbol{x}_{i, k}^{t}\|^{2}  \leq \frac{2 \eta^{2} \sigma^{2}\left(\tau -1\right)}{1-4 \eta^{2} L^{2}\tau \left(\tau -1\right)} \nonumber \\
&+\frac{4 \eta^{2}\tau\left(\tau -1\right)}{1-4 \eta^{2} L^{2}\tau \left(\tau -1\right)} \mathbb{E}\|\nabla \mathcal{L}_{i}(\boldsymbol{x}^{t})t\|^{2}.
\label{eq:bound on global and local models_2}
\end{align}

Similarly as in~\cite{wang2020tackling} where this analysis framework is proposed, we define $\gamma=4\eta^{2}L^{2}\tau \left(\tau -1\right) \leq 1$, and then \eqref{eq:bound on global and local models_2} can be simplified as,
\begin{align}
\frac{L^2}{\tau } \sum_{k=1}^{\tau} &\mathbb{E}\|\boldsymbol{x}^{t}\!\!-\!\!\boldsymbol{x}_{i, k}^{t}\|^{2}  \leq \frac{2 \eta^{2} \sigma^{2} L^{2}\left(\tau\!\! -\!\!1\right)}{1\!\!-\!\!\gamma} \!\!+\!\!\frac{\gamma}{1\!\!-\!\!\gamma} \mathbb{E}\|\nabla \mathcal{L}_{i}(\boldsymbol{x}^{t})\|^{2}.
\end{align}

Taking the average across all clients, there is
\begin{align}
    &\sum_{i=1}^{m} \frac{1}{m}\mathbb{E}\|\nabla \mathcal{L}_{i}(\boldsymbol{x}^{t})-\boldsymbol{h}_{i}^{(t)}\|^{2} \nonumber \\
    &\leq \!\frac{2 \eta^{2} \sigma^{2} L^{2}\left(\tau \!\! -\!\!1\right)}{1-\gamma} \!+\!\frac{\gamma}{1\!\!-\!\!\gamma}\sum_{i=1}^{m} \frac{1}{m} \mathbb{E}\|\nabla \mathcal{L}_{i}(\boldsymbol{x}^{t})\|^{2} \\
    & \leq \!\frac{2 \eta^{2} \sigma^{2} L^{2}\left(\tau\!\! -\!\!1\right)}{1\!\!-\!\!\gamma} \!\!+\!\frac{\gamma \beta^2}{1\!\!-\!\!\gamma} \mathbb{E}\|\nabla \mathcal{L}(\boldsymbol{x}^{t})\|^{2}\!\! +\! \frac{\gamma \zeta^2}{1\!\!-\!\!\gamma}.
    \label{eq: bound on global gradient and bound on normalized gradient}
\end{align}

\textbf{Global Convergence Property.} Based on the above analysis, we can bound the learning progress with \eqref{eq: bound on global gradient and bound on normalized gradient} and \eqref{eq:bound on learning progress_1},
\begin{align}
&\frac{1}{\tau \eta}\left(\mathbb{E}\left[\mathcal{L}(\boldsymbol{x}^{t+1})\right]-\mathcal{L}(\boldsymbol{x}^{t})\right) \nonumber \\
&\leq-\|\nabla \mathcal{L}(\boldsymbol{x}^{t})\|^{2}+\tau \eta L \sum_{i=1}^{m} \frac{1}{m^{2}}\left[1+3\left(1-p_{i}^{t}\right)^{2}\right] \sigma^{2} \nonumber \\
&+\frac{\gamma \beta^2}{2(1-\gamma)} \mathbb{E}\|\nabla \mathcal{L}(\boldsymbol{x}^{t})\|^{2} + \frac{\gamma \zeta^2}{2(1-\gamma)} \nonumber \\
&+\left(\frac{3}{2} \tau \eta L-\frac{1}{2}\right) \sum_{i=1}^{m} \frac{1}{m^{2}}\left(1-p_{i}^{t}\right)^{2} \mathbb{E}\|\boldsymbol{h}_{i^{\prime}}^{(t)}-\boldsymbol{h}_{i}^{(t)}\|^{2} \nonumber \\
& + \frac{ \eta^{2} \sigma^{2} L^{2}\left(\tau -1\right)}{1-\gamma}.
\label{eq:global_base}
\end{align}

If $\gamma \leq \frac{1}{2 \beta^{2}+1}$, then we have $\frac{1}{1-\gamma} \leq 1+\frac{1}{2 \beta^{2}}$ and $\frac{\gamma \beta^{2}}{1-\gamma} \leq \frac{1}{2}$. Therefore the above result can be simplified as,
\begin{align}
&\frac{1}{\tau \eta} \left( \mathbb{E}\left[\mathcal{L}(\boldsymbol{x}^{t+1})\right]-\mathcal{L}(\boldsymbol{x}^{t})\right) \nonumber \\
& \leq - \frac{3}{4}\|\nabla \mathcal{L}(\boldsymbol{x}^{t})\|^{2} + 4\tau \eta L \sum_{i=1}^{m} \frac{1}{m^{2}}\sigma^{2} \nonumber \\
&+ \eta^{2} \sigma^{2} L^{2}(\tau-1)\left ( 1+\frac{1}{2 \beta^{2}} \right ) \nonumber \\
&+\left(\frac{3}{2} \tau \eta L-\frac{1}{2}\right) \sum_{i=1}^{m} \frac{1}{m^{2}}\left(1-p_{i}^{t}\right)^{2} \mathbb{E}\|\boldsymbol{h}_{i^{\prime}}^{(t)}-\boldsymbol{h}_{i}^{(t)}\|^{2} \nonumber \\
& + \left [2 \eta^{2} L^{2} \tau(\tau-1)\left ( 1+\frac{1}{2 \beta^{2}} \right )\right ]\zeta^{2}.
\label{eq:beta}
\end{align}

Taking the average across all $T$ communication rounds,
\begin{align}
&\frac{1}{T} \sum_{t=0}^{T-1} \mathbb{E}\|\nabla \mathcal{L}(\boldsymbol{x}^{t})\|^{2} 
\leq  \frac{4\left[\mathcal{L}(\boldsymbol{x}^{0})-\mathcal{L}_{\mathrm{inf}}\right]}{3\eta \tau T} \nonumber \\
& + \frac{16\tau \eta L\sigma^{2}}{3m} +2 \eta^{2} \sigma^{2} L^{2}(\tau-1)  +  4 \eta^{2} L^{2} \tau(\tau-1)\zeta^{2} \nonumber \\
&+\left(2 \tau \eta L-\frac{2}{3}\right) \sum_{i=1}^{m} \frac{1}{m^{2}}\left(1-p_{i}^{t}\right)^{2} \mathbb{E}\|\boldsymbol{h}_{i^{\prime}}^{(t)}-\boldsymbol{h}_{i}^{(t)}\|^{2}. \nonumber
\end{align}

The proof of Lemma 2 is completed.

\subsection{Proof of Theorem 1}
From \eqref{eq:global_base}, due to the bounded similarity assumption on sparse models, we have
\begin{align}
&\frac{1}{\tau \eta}\left(\mathbb{E}\left[\mathcal{L}(\boldsymbol{x}^{t+1})\right]-\mathcal{L}(\boldsymbol{x}^{t})\right) \nonumber \\
&\leq-\|\nabla \mathcal{L}(\boldsymbol{x}^{t})\|^{2}+\tau \eta L \sum_{i=1}^{m} \frac{1}{m^{2}}\left[1+3\left(1-p_{i}^{t}\right)^{2}\right] \sigma^{2} \nonumber \\
& + \left(\frac{3}{2} \tau \eta L-\frac{1}{2}\right) \beta^2 \mathbb{E}\|\nabla \mathcal{L}(\boldsymbol{x}^{t})\|^{2} + \left(\frac{3}{2} \tau \eta L-\frac{1}{2}\right) \zeta^2 \nonumber \\
& + \frac{ \eta^{2} \sigma^{2} L^{2}\left(\tau -1\right)}{1-\gamma} \nonumber \\
&+\frac{\gamma \beta^2}{2(1-\gamma)} \mathbb{E}\|\nabla \mathcal{L}(\boldsymbol{x}^{t})\|^{2} + \frac{\gamma \zeta^2}{2(1-\gamma)}. 
\end{align}

With the above constraint on the learning rate, we have,
\begin{align}
&\frac{1}{\tau \eta}\left(\mathbb{E}\left[\mathcal{L}(\boldsymbol{x}^{t+1})\right]-\mathcal{L}(\boldsymbol{x}^{t}\right)) \nonumber \\
&\leq-\|\nabla \mathcal{L}(\boldsymbol{x}^{t})\|^{2}+4\tau \eta L \sum_{i=1}^{m} \frac{1}{m^{2}}\sigma^{2} \nonumber \\
&+\left[\frac{\gamma \beta^{2}}{2(1-\gamma)}+\frac{1}{4}\right] \mathbb{E}\|\nabla \mathcal{L}(\boldsymbol{x}^{t})\|^{2} + \left[\frac{\gamma }{2(1-\gamma)}+\frac{1}{4}\right] \zeta^2 \nonumber \\
& + \frac{ \eta^{2} \sigma^{2} L^{2}\left(\tau -1\right)}{1-\gamma}.
\end{align}

Similarly, with the same constraint on $\gamma$ and $\beta$ in~\eqref{eq:beta}, the above result can be simplified as,
\begin{align}
&\frac{1}{\tau \eta}\left(\mathbb{E}\left[\mathcal{L}(\boldsymbol{x}^{t+1})\right]-\mathcal{L}(\boldsymbol{x}^{t})\right) \nonumber \\
& \leq - \frac{1}{2}\|\nabla \mathcal{L}(\boldsymbol{x}^{t})\|^{2} + 4\tau \eta L \sum_{i=1}^{m} \frac{1}{m^{2}}\sigma^{2} \nonumber \\
&+ \eta^{2} \sigma^{2} L^{2}(\tau-1)\left ( 1+\frac{1}{2 \beta^{2}} \right ) \nonumber \\
& + \left [2 \eta^{2} L^{2} \tau(\tau-1)\left ( 1+\frac{1}{2 \beta^{2}} \right )+\frac{1}{4}\right ]\zeta^{2} \\ 
& \leq - \frac{1}{2}\|\nabla \mathcal{L}(\boldsymbol{x}^{t})\|^{2} + 4\tau \eta L \sum_{i=1}^{m} \frac{1}{m^{2}}\sigma^{2}\nonumber \\
&+ \frac{3}{2}\eta^{2} \sigma^{2} L^{2}(\tau-1)  + \left [3 \eta^{2} L^{2} \tau(\tau-1)+\frac{1}{4}\right ]\zeta^{2}. 
\end{align}

Taking the average across all rounds,
\begin{align}
&\frac{1}{T} \sum_{t=0}^{T-1} \mathbb{E}\|\nabla \mathcal{L}(\boldsymbol{x}^{t})\|^{2}
\leq  \frac{2\left[\mathcal{L}(\boldsymbol{x}^{0})-\mathcal{L}_{\mathrm{inf}}\right]}{\eta \tau T} \\
& + \frac{8\tau \eta L\sigma^{2}}{m} +3 \eta^{2} \sigma^{2} L^{2}(\tau-1)  + \left [6 \eta^{2} L^{2} \tau(\tau-1)+\frac{1}{2}\right ]\zeta^{2}. \nonumber
\end{align}

For the ease of writing, we define $A = \tau$, $B = \tau-1$ and $C = \tau(\tau-1)$, and then we derive
\begin{align}
&\frac{1}{T} \sum_{t=0}^{T-1} \mathbb{E}\|\nabla \mathcal{L}(\boldsymbol{x}^{t})\|^{2} 
\leq  \frac{2\left[\mathcal{L}(\boldsymbol{x}^{0})-\mathcal{L}_{\mathrm{inf}}\right]}{\eta \tau T} \nonumber \\
& + \frac{8 \eta L\sigma^{2}A}{m} +3 \eta^{2} \sigma^{2} L^{2}B  + \left(6 \eta^{2} L^{2} C +\frac{1}{2}\right)\zeta^{2}.
\end{align}

Since there is
\begin{align}
    \min_{t \in \mathbb{T}} \mathbb{E}\!\|\nabla \mathcal{L}(\boldsymbol{x}^{t})\|^{2}  \! \leq  \! \frac{1}{T}\! \sum_{t=0}^{T-1} \mathbb{E}\!\|\nabla \mathcal{L}(\boldsymbol{x}^{t})\|^{2},
\end{align}
it holds that,
\begin{align}
&\min _{t \in \mathbb{T}} \mathbb{E}\|\nabla \mathcal{L}(\boldsymbol{x}^{t})\|^{2}
\leq  \frac{2\left[\mathcal{L}(\boldsymbol{x}^{0})-\mathcal{L}_{\mathrm{inf}}\right]}{\eta \tau T} \nonumber \\
& + \frac{8 \eta L\sigma^{2}A}{m} +3 \eta^{2} \sigma^{2} L^{2}B  + \left(6 \eta^{2} L^{2} C +\frac{1}{2}\right)\zeta^{2}.
\end{align}

By setting $\eta=\sqrt{\frac{m}{\tau T}}$, we have
\begin{align}
    \min _{t \in \mathbb{T}} &\mathbb{E}\|\nabla \mathcal{L}(\boldsymbol{x}^{t})\|^{2} \leq \mathcal{O}\left(\frac{1}{\sqrt{m \tau T}}\right)+\mathcal{O}\left(\frac{A \sigma^{2}}{\sqrt{m \tau T}}\right) \nonumber \\
    &+\mathcal{O}\left(\frac{m B \sigma^{2}}{\tau T}\right)\!+\!\mathcal{O}\left(\frac{m C \zeta^{2}}{\tau T}\right). \nonumber
\end{align}

The proof of Theorem 1 is completed.


\bibliography{reference}
\bibliographystyle{IEEEtran}


 




\end{document}